\documentclass[reprint, superscriptaddress, amsmath,amssymb, aps, prresearch, longbibliography, floatfix]{revtex4-2}

\usepackage{amsmath}
\usepackage{graphicx}
\usepackage{ulem}
\usepackage{bm}
\usepackage{natbib}
\usepackage{dsfont}
\usepackage{tikz}
\usepackage{epsfig}
\usepackage{feynmf}
\usepackage{blindtext, rotating}
\usepackage{mathtools}
\usepackage{dsfont}
\usepackage{subcaption}
\usepackage{physics}
\usepackage{amsfonts}
\usepackage{xcolor}
\usepackage{ragged2e}
\usepackage{siunitx}
\usepackage{comment}
\usepackage{soul}
\usepackage{lipsum}
\usepackage{hyperref} 
\hypersetup{breaklinks=true, colorlinks=true, citecolor=blue, linkcolor=cyan, urlcolor=blue,filecolor=blue}

\usepackage{xcolor} 

\DeclareCaptionJustification{justified}{\justifying}

\DeclareSIUnit{\rad}{rad}

\definecolor{bright_blue}{HTML}{85C1E9}
\definecolor{middle_blue}{HTML}{2E86C1}
\definecolor{dark_blue}{HTML}{1B4F72}

\begin{document}

\title{Quantum-induced Stochastic Optomechanical Dynamics}





\author{Pedro V. Paraguass\'u}
\affiliation{Department of Physics, Pontifical Catholic University of Rio de Janeiro, Rio de Janeiro 22451-900, Brazil}

\author{Luca Abrah\~ao}
\affiliation{Department of Physics, Pontifical Catholic University of Rio de Janeiro, Rio de Janeiro 22451-900, Brazil}

\author{Thiago Guerreiro}
\email{barbosa@puc-rio.br}
\affiliation{Department of Physics, Pontifical Catholic University of Rio de Janeiro, Rio de Janeiro 22451-900, Brazil}

\begin{abstract}

We study the effective stochastic dynamics of a semiclassical probe induced by linear optomechanical interactions with a quantum oscillator. To do so, we introduce path integrals and the method of Feynman-Vernon influence functionals in quantum optics and analyse the semiclassical dynamics of a levitated nanoparticle interacting with quantum light, as well as with another quantum particle. In all cases, quantum fluctuations ubiquitously lead to state-dependent non-equilibrium noise. Notably, this noise can be exponentially enhanced by wavepacket delocalization, i.e. quantum squeezing, and displays both a stationary and a non-stationary contribution with intricate dependence on the squeezing angle. For the case of nanoparticles coupled by the Coulomb interaction such noise can imprint potentially measurable signatures in multiparticle levitation experiments. We also discuss the case in which the mechanical oscillators are coupled by gravity, and the relation of the quantum-induced noise to gravitational-induced entanglement. Quantum-induced optomechanical fluctuations also hold strong analogy to quantum gravitational wave noise and interconnect stochastic thermodynamics, graviton physics and the detection of gravity-mediated entanglement.
\end{abstract}


\maketitle

\section{Introduction}

Quantum theory does not predict its own limit of validity. In principle, a system of any given mass can exhibit quantum behavior, as beautifully displayed by matter-wave interference experiments of molecules with increasing numbers of atoms \cite{fein2019quantum, Brand2020}. As the system's size grows, however, so does its coupling to the environment. This typically leads to decoherence and the emergence of classicality, making it in practice very hard to witness quantum effects with macroscopic massive objects \cite{zurek2003decoherence}. To overcome this challenge, optically levitated nanoparticles have surfaced as a promising system \cite{millen2020optomechanics, gonzalez2021levitodynamics}. Levitation provides extreme isolation and a high degree of control \cite{romero2011large}, enabling ground state cooling in cavity \cite{delic2019cavity, delic2020cooling} and free space environments \cite{magrini2021real, tebbenjohanns2021quantum}, multiparticle interactions via optical and electrostatic mechanisms \cite{rieser2022tunable, penny2023sympathetic, livska2023cold, vijayan2023cavity}, hybrid and compact trapping \cite{bykov2022hybrid, bonvin2023hybrid, melo2023vacuum}, strong light-matter coupling \cite{de2021strong, dare2023linear} and large expansion of the motional mechanical state \cite{bonvin2023state, mufato2023}. Altogether, these developments motivate the main question of our work: can we use optically levitated nanoparticles as a platform to study the interactions between a mesoscopic massive quantum system and a classical probe? 
%


The question of quantum-classical (QC) interactions is at the core of fundamental physics. 
Intuitively, quantum fluctuations induce noise in an interacting classical system, producing an effective stochastic dynamics. 
As we will see, this noise explicitly depends on Planck's constant ($ \hbar $) and its origin can be traced down to the entanglement formed between the probe and the unobserved quantum system before taking the semiclassical approximation. Due to the existence of a ground state and the quantization of energy, we can expect this quantum-induced noise to be out-of-equilibrium \cite{clerk2010introduction, milburn2011introduction}. Hence the study of quantum-induced noise can be investigated from the viewpoint of stochastic thermodynamics \cite{sekimoto2010stochastic, seifert2012stochastic, peliti2021stochastic}, extending mesoscopic non-equilibrium optomechanical experiments \cite{brunelli2015out, dechant2015all, debiossac2020thermodynamics, debiossac2020thermodynamics} to the realm of quantum-induced stochastic processes. 

\begin{figure}[ht!]
    \centering
    \includegraphics[width=0.48\textwidth]{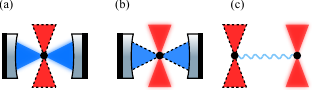}
    \caption{Different quantum-classical optomechanical interactions. (a) Quantum light influencing a classical particle. (b) Quantum particle influencing classical light. (c) Quantum particle influencing a classical particle.}
    \label{fig:setup}
\end{figure}

Moreover, QC interactions play a central role in graviton physics \cite{hu2008stochastic, coradeschi2021can} and tabletop quantum gravity proposals \cite{bose2017spin, carney2019tabletop, carlesso2019testing, aspelmeyer2022zeh}. On one hand, we currently have no direct experimental evidence for the quantization of gravity and even the need for a quantum description of gravitational phenomena has been questioned in a number of different contexts \cite{penrose1996gravity, jacobson1995thermodynamics, dyson2013graviton, carney2023graviton, tobar2023detecting, oppenheim2023postquantum}.
On the other, it has been recently suggested within the effective field theory description of gravity that quantum gravitational waves (GWs) can imprint detectable signatures in interferometers \cite{guerreiro2020quantum, parikh2021quantum, parikh2021signatures, cho2022quantum}, while delocalized quantum particles can become entangled solely by the action of the gravitational interaction \cite{bose2017spin, aspelmeyer2022zeh}. Detection of non-classical gravitational noise would provide a stepping stone to the debate on the need for a quantum theory of gravity. However, given the weakness of gravity, measuring gravitational noise is a daunting objective and experimentally probing analogous effects in levitated optomechanics might lead to new paths forward in that direction \cite{guerreiro2022quantum}.  In addition, measuring the stochastic dynamics of a classical test particle induced by the Coulomb field of a highly delocalized quantum particle is warm up to the more difficult task of measuring the effects of electromagnetic and gravitational-mediated entanglement \cite{rudolph2022force, feynman2018feynman, anastopoulos2015probing, belenchia2018quantum, bengyat2023gravity}. We highlight that, while it is a possibility that the gravitational field is inherently classical \cite{penrose1996gravity, oppenheim2023postquantum}, the notion of classicality in quantum optomechanics arises due to decoherence and interaction with the environment \cite{zurek2003decoherence}.

Here we investigate three ways in which levitated nanoparticles might be used to probe QC dynamics and quantum-induced noise. First, we consider a classical nanoparticle interacting with a quantum mechanical cavity via the linear optomechanical coupling, as conceptualized in Fig. \ref{fig:setup}(a). We take coherent scattering (CS) as the prototypical light-matter interaction in levitated optomechanics \cite{delic2019cavity}. Since the linear optomechanical interaction is symmetric with respect to the two modes, we can extend our investigation to the case of a classical optical mode interacting with a quantum particle, as depicted in Fig \ref{fig:setup}(b). We then analyse the case of a quantum and classical particles coupled via the linearised Coulomb interaction as in Fig. \ref{fig:setup}(c). At this point, we can use the analogy between the Coulomb and Newtonian potentials to comment on the gravitational field of a delocalized source. In all three cases, we derive the effective stochastic semiclassical dynamics induced by the quantum system using the formalism of double path integrals \cite{feynman1965path} and the Feynman-Vernon influence functional \cite{FEYNMAN1963118, caldeira1983path}. 
We note that, although the use of path integrals is not strictly necessary -- other methods such as quantum Langevin equations can be used to obtain equivalent results for linear systems -- the use of influence functionals brings new physical insights. Notably, when compared to the canonical approach, path integrals greatly simplify perturbative calculations and clarify the meaning of the semiclassical approximation. Furthermore, the same methods can be used to generalize perturbative calculation to nonlinear interacting systems such as the dispersive interaction \cite{aspelmeyer2014cavity}.

Following recent work on quantum GWs \cite{guerreiro2020quantum, parikh2021quantum, parikh2021signatures}, we study how the effective stochastic dynamics is affected by the state of the quantum system, in particular the ground, squeezed-coherent and squeezed-thermal states, all achievable in laboratory experiments. Despite the interaction of a GW with a detector being governed by a cubic derivative Hamiltonian \cite{parikh2021signatures}, in principle very different from the linear optomechanical coupling, we find strong similarities between both cases. This reinforces the analogy between optomechanics and quantum GWs \cite{pang2018quantum, guerreiro2020quantum, guerreiro2022quantum}. Likewise in the case of GWs \cite{parikh2021quantum}, squeezing of the quantum state exponentially enhances the quantum-induced noise and adds a non-stationary noise component. While we see that quantum-induced noise is very small for the linear light-matter interactions achieved in current levitated optomechanics experiments, a delocalized quantum particle might imprint significant fluctuations in the position of a semiclassical particle interacting via the Coulomb potential.
Combined with efforts to delocalize the wavefunction of a nanoparticle \cite{weiss2021large, kustura2022mechanical, neumeier2022fast}, quantum-induced stochastic dynamics could be significantly enhanced to an observable level in near-future multiparticle experiments. 

This paper is organized as follows. In Sec. \ref{FVformalism} we import double path integrals and the method of Feynman-Vernon influence functionals to the context of linear optomechanical systems. We show how to calculate the influence functional and stochastic path integral propagator for different quantum states in the case of a single mode field, notably the vacuum, squeezed-coherent and squeezed thermal states. In Sec. \ref{trapedparticle} we extend these methods to sum over the modes of an optical cavity and treat the problem of a trapped nanoparticle as a semiclassical probe of quantum light. Sec. \ref{sec:light-particle} deals with the converse problem, of a semiclassical cavity as a probe of a quantum particle, while \ref{sec:particle-particle} treats the case of a classical and quantum particles interacting via the Coulomb potential. The reader who might wish to skip the technical details concerning path integrals and focus on possible experimental implications and a discussion of the analogous gravitational effects can go directly to Sec. \ref{sec:particle-particle}. Finally, in section \ref{sec:dissucssion}, we conclude with a brief discussion of the results and directions for future work.

\section{Path Integral formulation of linear optomechanical system}\label{FVformalism}

Path integrals fundamentally deal with fluctuations, classical or quantum. We are interested in deriving a classical stochastic equation where fluctuations arise from interactions with a quantum system. Therefore, double path integrals for density matrices and the Feynman-Vernon method are the natural tools. In order to bridge the gap between the Feynman-Vernon method and quantum optics and optomechanics, we have adapted the theory by combining coherent state and configuration space path integrals in a single expression for the density matrix. 
We find it convenient to express the dynamics of the semiclassical system in terms of variables in configuration space (position and velocity) while treating the quantum system in terms of the coherent state basis.


\subsection{Hamiltonian}

We consider a linear optomechanical system consisting of a levitated nanoparticle of mass $ m $ and frequency $ \omega_{m} $ coupled to an optical cavity mode with frequency $ \omega $. 
The total Hamiltonian of the system reads
\begin{eqnarray}
    H = H_{c} + H_{m} + H_{I}
    \label{total_hamiltonian}
\end{eqnarray}
where the free cavity and mechanical Hamiltonians are
\begin{eqnarray}
    H_{c}/\hbar &=&   \omega\, a^\dagger a \label{optical_hamiltonian} \\
    H_{m}/\hbar &=& \omega_m b^\dagger b
    \label{mechanical_hamiltonian}
\end{eqnarray}
and the interaction reads
\begin{eqnarray}
    H_{I}/\hbar = g  q(t) X(t)
    \label{interaction_hamiltonian}
\end{eqnarray}
Here, $ q(t)$ and $ X(t)$ denote the dimensionless mechanical position and cavity mode amplitude quadratures, respectively. In terms of creation and annihilation operators these quantities are $ q = b + b^{\dagger} $ and $ X = a + a^{\dagger}  $. To recover the dimensionful position we multiply by the zero point fluctuation,
\begin{eqnarray}
    \mathbf{q} &\equiv& q_{0} q = q_{0} (b + b^{\dagger})
\end{eqnarray}
where $ q_{0} = \sqrt{\hbar / 2m\omega_{m}} $.



Following \cite{FEYNMAN1963118, parikh2021signatures, parikh2021quantum, cho2022quantum}, we are interested in studying how state-dependent quantum fluctuations of one oscillator affects the semiclassical dynamics of the other via the interaction $ H_{I}$. To this end, we will construct a path integral expression for the system's density matrix and employ the theory of Feynman-Vernon influence functionals. In doing this, we introduce coherent state path integrals, a convenient resource for when the Hamiltonian is expressed in terms of bosonic creation and annihilation operators \cite{hillery1982path}. 

\subsection{Density matrix}

We start with the reduced mechanical density matrix  in the position basis,
\begin{equation}
  \rho_{b}(q_{t},q'_{t},t) = \frac{1}{\pi}\int{{d^2\beta}} \bra{q,\beta} U \rho_{0}U^{\dagger}\ket{q',\beta}
\end{equation}
where $ \ket{q',\beta} \equiv \vert q' \rangle \otimes \vert \beta \rangle $ is an element of the position-coherent state product bases with a similar definition for its dual, $ \beta $ is the coherent state amplitude variable, $ U $ is the unitary evolution of the system generated by \eqref{total_hamiltonian} and $ \rho_{0} $ is the initial optomechanical joint density matrix, assumed to be a product state $ \rho_{0} = \rho_{0,a} \otimes \rho_{0,b} $, which is justified provided each subsystem is prepared independently and the interaction Hamiltonian \eqref{interaction_hamiltonian} is switched on over a time scale much shorter than the inverse characteristic frequencies $ \omega_{m}^{-1}, \omega^{-1} $. We will denote $ q_{t} \equiv q(t) $ the \textit{forward} position variable, while $ q'_{t} \equiv q'(t) $ will be referred to as the \textit{backward} position variable. 
Inserting complete bases 
$ \frac{1}{\pi}\int dq_{0} d^{2}\beta_{0} \vert q_{0}, \beta_{0} \rangle \langle q_{0}, \beta_{0} \vert  $
of joint position-coherent state states we can rewrite the reduced density matrix in terms of coherent state propagators,
\begin{eqnarray}
    \rho_{b}(q_{t},q'_{t},t) = \dfrac{1}{\pi^{3}} \int d^{2}\beta dq_{0}dq'_{0} d^{2}\beta_{0} d^{2}\beta'_{0} \  \nonumber \\
    \times K(q_{t}, \beta ; q_{0}, \beta_{0}) \  \langle q_{0}, \beta_{0} \vert \rho_{0} \vert q'_{0}, \beta'_{0} \rangle \ K^{*}(q'_{t}, \beta ; q'_{0}, \beta'_{0})
\end{eqnarray}

\noindent where,
\begin{eqnarray}
    K(q_{t}, \beta ; q_{0}, \beta_{0}) &=& \bra{q_{t},\beta} U \ket{q_{0},\beta_{0}}  \\
    K^{*}(q'_{t}, \beta ; q'_{0}, \beta'_{0}) &=& \bra{q'_{0},\beta'_{0}} U^{\dagger} \ket{q'_{t},\beta}
\end{eqnarray}
These propagators can in turn be written as the path integral \cite{hillery1982path};
\begin{eqnarray}
    K(q_t,\beta|q_0,\beta_0) =\int_{q_0}^{q_t} \int_{\beta_0}^{\beta}\mathcal{D}q  \mathcal{D}\alpha \nonumber \\
    \times e^{\int_{0}^{t_f} dt \left[\frac{1}{2}\left(\alpha\dot\alpha^*-\alpha^*\dot\alpha\right)-\frac{i}{\hbar}\left(H_{c}(\alpha,\alpha^*)+ H_{I}(q,\alpha, \alpha^{*}) -L_{\mathbf{q}}\right)\right]}, \label{propagator_coherent}
\end{eqnarray}
where $ H_{c}(\alpha,\alpha^*) $ and $ H_{I}(q,\alpha, \alpha^{*}) $ denote the optical and interaction Hamiltonians in Eqs. \eqref{optical_hamiltonian} and \eqref{interaction_hamiltonian} with the substitutions $ a \rightarrow \alpha, a^{\dagger} \rightarrow \alpha^{*} $, and $ L_{\mathbf{q}} $ is the Lagrangian of the mechanical oscillator, 
\begin{eqnarray}
    L_{\mathbf{q}} = \frac{1}{2}m \left( \dot{\mathbf{q}}^{2} - \omega_{m}^{2} \mathbf{q}^{2} \right) 
\end{eqnarray}
The conjugate propagator $ K^{*}(q'_{t}, \beta ; q'_{0}, \beta'_{0}) $ is obtained from Eq. \eqref{propagator_coherent} by complex conjugation. 

For a separable initial joint state $
\langle q_{0}, \beta_{0} \vert \rho_{0} \vert q'_{0}, \beta'_{0} \rangle = \rho_{0,a}(\beta_{0}, \beta'_{0}) \rho_{0,b}(q_{0}, q'_{0}) $ and we may write the reduced mechanical density matrix as
\begin{equation}
    \rho_{b}(q_t,q'_t,t) = \int dq_0 dq'_0 \,\rho_{0,b}(q_0,q'_0) \mathcal{J}(q_t,q'_t|q_0,q'_0), 
\end{equation}
where we define the \textit{double path integral propagator}, 
\begin{align}
     \mathcal{J}(q_t,q'_t|q_0,q'_{0})  = \int_{q_0,q'_{0}}^{q_t,q'_t} \mathcal{D}q\mathcal{D}q'
     e^{\frac{i}{\hbar}\int_{0}^{t_f} dt \left(L_{\mathbf{q}}-L_{\mathbf{q}'}\right)}\mathcal{F}[q,q'], \label{bigaction}
\end{align}
and $\mathcal{F}[q,q']$ denotes the Feynman-Vernon \textit{influence functional} \cite{feynman2010quantum},
    \begin{align}
    \mathcal{F}[q,q'] = \frac{1}{\pi^3}\int d^2\beta  d^2\beta_0 d^2\beta'_0 \;\rho_{0,a}(\beta_0,\beta'_0) \nonumber \\
    \times \int_{\beta_0}^{\beta} \int_{\beta'_{0}}^{\beta'} \mathcal{D}\alpha  \mathcal{D}\alpha' \;e^{\int_{0}^{t_f} dt \left[\frac{1}{2}\left(\alpha\dot\alpha^*+\alpha'\dot\alpha'^*-\alpha^*\dot\alpha-\alpha'^*\dot \alpha'\right)\right]}
    \nonumber\\ 
    \times e^{\int_{0}^{t_f}dt\left[-\frac{i}{\hbar}(H_c(\alpha,\alpha^*)-H_c(\alpha',\alpha'^*)+H_I(q,\alpha)-H_I(q',\alpha'))\right]}.\label{influence}
\end{align}
Since we are dealing with the density matrix, i.e. probabilities rather than amplitudes, we have a double path integral. Furthermore, the path integral treats the mechanical oscillator in the position bases while the cavity mode is dealt with in the coherent state bases.  Moreover, we can see that both the Lagrangian and Hamiltonian make an appearance in the expression of the stochastic propagator, reminiscent of Routhian mechanics \cite{lemos2018analytical}. 

Since the interaction Hamiltonian is symmetric with respect to the cavity and mechanical modes, the same derivation can be carried over with the mechanical system in the coherent state bases, which will be convenient for when we want to study the effective dynamics of the optical mode influenced by the mechanical oscillator. 

The influence functional can be evaluated using the canonical formalism. 
Define the time-ordered evolution operator,
\begin{eqnarray}
    U_{q} = \mathcal{T} \exp\left(-\frac{i}{\hbar}\int_{0}^{t_f} dt\,  {H}_{q}[q(t)] \right),
\end{eqnarray}
and the reduced propagators,
\begin{eqnarray}
    K_{q}(\beta ; \beta_{0}) &=&\bra{\beta}  {U}_q\ket{\beta_0} \\
    K_{q}^{\dagger}(\beta ; \beta_0') &=& \bra{\beta_{0}}  {U}_q^{\dagger}\ket{\beta}
\end{eqnarray}
where 
\begin{eqnarray}
    H_{q}[q(t)] = H_{c} + H_{I}
\end{eqnarray}
is a restricted cavity-interaction Hamiltonian.
For an initial pure state $ \rho_{0,a} = \ket{\psi}\bra{\psi} $ the influence functional becomes
\begin{eqnarray}
      \mathcal{F}[q,q'] &=&  \frac{1}{\pi^3}\int d^2\beta  d^2\beta_0  d^2\beta_0'  \rho_{0,a}(\beta_{0},\beta_0') \nonumber \\
      &\times& K_{q}(\beta ; \beta_0) K_{q}^{\dagger}(\beta ; \beta_0') \nonumber \\ 
    &=& \frac{1}{\pi}\int d^2\beta \bra{\beta} U_q \ket{\psi}\bra{\psi}U_{q'}^{\dagger}\ket{\beta} \nonumber  \nonumber \\&=& \bra{\psi}U^{ \dagger}_{q'}U_{q}\ket{\psi}. \label{influ}
\end{eqnarray}
Note that this expression can be generalized to mixed states by using the trace formula for the expectation value. 
 {Moreover, since there are no interactions at $t=0$, we have that $\mathcal{F}[q,q']\big|_{t=0}=1$}
We can move to the interaction picture via the transformation $U^\dagger_q\rightarrow U^{I\dagger}_{q'}=U^\dagger_{q'}e^{-\frac{i}{\hbar}H_c t}$, where without loss we choose $t_0=0$.  The influence functional is then given by
\begin{equation}
    \mathcal{F}[q,q']=\bra{\psi}U^{ I\dagger}_{q'}U^I_{q}\ket{\psi} \ ,\label{cruinfluence}
\end{equation}
where we define the interaction picture evolution
\begin{equation}
    U^I_{q} = \mathcal{T} \exp\left(-\frac{i}{\hbar}\int_{0}^{t_f} dt\,  {H_I}[q(t)] \right)
\end{equation}
When the commutator of the interaction Hamiltonian is a c-number the time ordered symbol $ \mathcal{T} $ can be exchanged by a commutator term \cite{itzykson2012quantum}, 
\begin{eqnarray}
     {U}_{q}^I = e^{-\frac{i}{\hbar}\int_0^{\tau} dt\, H_I} e^{-\frac{1}{2\hbar^2}\int_0^{t_f}\int_0^{t} dt\,dt'\left[ H_I(t), H_I(t')\right]} \label{itzukson}
\end{eqnarray}
\noindent with
\begin{equation}
    H_I/\hbar = g\, q(t) \left(a^\dagger(t)+a(t)\right),
\end{equation}
and
\begin{eqnarray}
    a^\dagger(t) = {a}^\dagger e^{i\omega t},\;\; a(t)={a} e^{-i\omega t}.
\end{eqnarray}

After evaluation of the influence functional all the dependence on the optical degrees of freedom is removed, leaving only path integrals in the forward and backward position variables 
$q$ and $q'$.

\subsection{Influence functional}

As observed in \cite{parikh2021signatures}, by manipulating Eq.~(\ref{cruinfluence}) via the Baker–Campbell–Hausdorff formula, we can further simplify the influence functional to
\begin{equation}
    \mathcal{F}[q,q'] = e^{i\Phi_{0}[q,q']} F_\psi[q,q'] \label{braketpart}
\end{equation}
where the \textit{influence phase} $ \Phi_{0}[q,q'] $ splits into two parts,
\begin{eqnarray}
    \Phi_{0}[q,q'] = i\Phi_0^{\mathrm{fl}}+i\Phi_0^{\mathrm{diss}}
    \label{vacuumphase}
\end{eqnarray}
one corresponding to fluctuations,
\begin{eqnarray}
    i\Phi_0^{\mathrm{fl}} = - \frac{g^{2}}{2}\int_0^{t_f}\int_0^{t_f}dt\, dt' \nonumber \\
    \times \left(q(t)-q'(t)\right)\left(q(t')-q'(t') \right)\cos\left(\omega(t-t')\right)
\end{eqnarray}
and another to dissipation
\begin{eqnarray}
    i\Phi_0^{\mathrm{diss}} =  ig^{2} \int_0^{t_f}\int_0^{t}dt\,dt' \nonumber \\
    \times \left(q(t)-q'(t)\right)\left(q(t')+q'(t')\right)\sin\left(\omega(t-t')\right) 
    \label{dissipation_phase_vacuum}
\end{eqnarray}
Further, Eq. \eqref{braketpart} has the state-dependent factor
\begin{eqnarray}
    F_\psi[q,q'] = \bra{\psi}e^{-a^\dagger W^*}e^{a W} \ket{\psi},
\end{eqnarray}
where $ W $ is
\begin{equation}
     W = - ig \int_0^{t_f} dt\;(q(t)-q'(t))e^{-i\omega t}.
\end{equation}
Note that $ F_\psi[q,q'] $ is the quantum optical characteristic function evaluated at $ W $ \cite{milburn2012quantum}. The influence phase $ \Phi_{0}[q,q'] $ encodes the effects of vacuum quantum fluctuations of the optical mode upon the mechanical oscillator and is present for all quantum states of the optical field. In general, the vacuum influence will be modified by the state-dependent term $ F_\psi[q,q'] $.
We next discuss each of these terms in detail, starting from a calculation of the effects associated to the vacuum influence phase.
We then generalize the results to squeezed-coherent and thermal states. We note that the Gaussian character of these states significantly simplifies the problem at hand, but in principle nothing prevents one from extending the path integral method to include the effects associated to non-Gaussian states \cite{lawande1995path}.

\subsection{Fluctuation and Dissipation}
The fluctuation contribution to the influence phase can be written as
   \begin{eqnarray}
   i\Phi_0^{\mathrm{fl}} = - \frac{1}{2} \int_0^{t_f}\int_0^{t_f} dt\, dt' J(t) \mathcal{A}(t,t') J(t')
   \label{fluctuation_phase}
\end{eqnarray} 
where we define
\begin{eqnarray}
    J(t) = q(t)-q'(t)
\end{eqnarray}
and the two-time noise kernel
\begin{eqnarray}
    \mathcal{A}(t,t') = g^2\cos(\omega(t-t'))
\end{eqnarray}

\noindent  {We now perform the so-called Feynman-Vernon trick \cite{feynman2010quantum}, or Hubbard-Stratonovich transformation \cite{mccaul2021win}}, which consists in the observation that Eq. \eqref{fluctuation_phase} can be written in terms of a path integral over an auxiliary variable $\zeta(t)$,
\begin{align}
    e^{-\frac{1}{2}\int_0^{t_f}\int_0^{t_f}dt\;dt' J(t)\mathcal{A}(t,t')J(t')} = \nonumber \\
    \int \mathcal{D}\zeta e^{{-\frac{1}{2}\int_0^{t_f}\int_0^{t_f} dt\,dt' \zeta(t)\mathcal{A}^{-1}(t,t')\zeta(t')} + i \int_0^{t_f} dt \zeta(t)J(t)}\label{feytrick}
\end{align}
where $ \int_{0}^{t_f} ds \mathcal{A}(t,s)\mathcal{A}^{-1}(s,t') = \delta(t-t') $. Effectively, the Feynman trick decouples the forward and backward variables $q$ and $q'$ at the expense of introducing the variable $\zeta$, which is interpreted as a random process with the probability density functional of the stochastic variable,
\begin{equation}
    P[\zeta(t)] = \exp\left({-\frac{1}{2}\int_0^{t_f}\int_0^{t_f} dt\,dt' \zeta(t)\mathcal{A}^{-1}(t,t')\zeta(t')}\right).\label{density}
\end{equation}
This random process can be thought of as the noise induced by quantum fluctuations of the optical mode on the mechanical oscillator.

Denoting the stochastic average over $ P[\zeta(t)] $ by $\langle \dots \rangle_\zeta$, Eq.~\eqref{feytrick} can be written as, 
\begin{eqnarray}
    \langle e^{\frac{i}{\hbar}\int_0^{t_f}dt \zeta(t)J(t)} \rangle_\zeta \ .
\end{eqnarray}

\noindent The mean value and time correlation function of this noise are,
\begin{eqnarray}
    \langle \zeta(t)\rangle_\zeta &=& 0, \\
    \langle \zeta(t)\zeta(t')\rangle_\zeta &=& \mathcal{A}(t,t'),
\end{eqnarray}
Since the stochastic variable is Gaussian, the first and second moments are sufficient to completely describe the process. Note that $ \mathcal{A}(t,t') \propto g^{2} $, which implies $ \zeta(t) \propto g $. It will be convenient to write
\begin{eqnarray}
    \zeta(t) = g \xi(t) \ .\label{dimensionnoise}
\end{eqnarray}
where we define the \textit{dimensionless random variable} $ \xi(t) $.


We now turn to the dissipation contribution given in Eq. \eqref{dissipation_phase_vacuum}.
This term is non-local in time \cite{ferialdi2012functional, heredia2022nonlocal} and is given by the integral of a product of two distinct functions, namely $ J(t) = (q(t) - q'(t)) $ and $ (q(t) + q'(t)) $. Unlike $ i\Phi_0^{\mathrm{fl}} $, the Feynman trick cannot be performed and it cannot be associated to fluctuations.
Consequently, the equations of motion for the forward and backward variables $ q $ and $ q' $ are coupled \cite{ferialdi2012functional,heredia2022nonlocal}. This coupling is associated to the breaking of time-reversal symmetry \cite{debiossac2020thermodynamics}, but as we will see, its effects are subleading when compared to the fluctuations. 



\subsection{Equation of motion \label{sec:eq_motion}}


For the ground state $ \ket{\psi} = \ket{0} $, the total influence phase yields the propagator,
\begin{widetext}
    \begin{eqnarray}
        \mathcal{J}(q_t,q'_t|q_0,q'_{0})  =\int_{q_0,q'_{0}}^{q_t,q'_t} \mathcal{D}q\mathcal{D}q' \int \mathcal{D}\zeta  \exp\left({-\frac{1}{2}\int_0^{t_f}\int_0^{t_f} dtdt' \zeta(t)\mathcal{A}^{-1}(t,t')\zeta(t')}\right) 
        \exp\left(\frac{i}{\hbar}\int_{0}^{t_f} dt \left(L_{\mathbf{q}} - L_{\mathbf{q}'}\right)\right)\nonumber \\\exp\left(\frac{i}{\hbar}\int_0^{t_f} dt \;
        \left( \frac{\hbar \zeta(t)}{q_{0}} \right)(\mathbf{q}(t)-\mathbf{q}'(t))+ \frac{i}{\hbar} \left(\frac{\hbar g^{2}}{q_{0}^{2}} \right) \int_0^{t_f}\int_0^{t}dt\, dt'\left(\mathbf{q}(t)-\mathbf{q}'(t)\right)\left(\mathbf{q}(t')+\mathbf{q}'(t')\right)\sin\left(\omega(t-t')\right)\right)
        \label{final__mechanical_propagator}
    \end{eqnarray}
\end{widetext}
The first line shows the stochastic density kernel and the dynamics of the mechanical oscillator. The second line contains the coupling between the stochastic variable $ \zeta(t) $ and the forward and backward position variables. Note here that writing the arguments in the exponential in terms of the dimensionful oscillator position forces the appearance of the combination $ \hbar \zeta / q_{0} = \hbar g \xi / q_{0} $, which has dimension of force. Finally, the last term consists in the non-local dissipation coupling the $ \mathbf{q} $ and $ \mathbf{q}' $ variables.  {The above expression highlights the advantage of using path integrals, where the Lagrangian is expanded with all the involved physical terms, providing physical insight into the semi-classical dynamics of the system, our primary interest.}

Extremising the argument in the exponentials of \eqref{final__mechanical_propagator} with respect to $ \mathbf{q}(t)$ and $\mathbf{q}'(t)$ leads to effective Langevin-like equations for the forward and backward paths. In general these are coupled stochastic differential equations, with the coupled terms arising from the cross-terms in the dissipation phase $ \Phi_0^{\mathrm{diss}}  $. As customary, we will neglect the coupling between forward and backward paths and take the ansatz $ \mathbf{q}(t) = \mathbf{q}'(t)$ \cite{parikh2021signatures, kamenev2023field}. As consequence, the dynamics of the forward and backward paths obey the same equation of motion. This corresponds to expressing the action in terms of symmetric $(q(t) + q'(t))/2$ and antisymmetric $(q(t) - q'(t))$ paths and expanding to leading order terms in the anti-symmetric path, as was also done in \cite{parikh2021signatures}. 

We find the Langevin-like equation of motion
\begin{equation}
    m \ddot{\mathbf{q}}(t) + m\omega_{m}^{2} \mathbf{q} = \mathbf{f}_{\mathcal{Q}}(t) + \mathbf{f}_{\rm diss}(t) \label{langevin}
\end{equation}
where we define the \textit{quantum fluctuation force},
\begin{eqnarray}
    \mathbf{f}_{\mathcal{Q}}(t) \equiv \frac{\hbar \zeta(t)}{q_{0}} = \left(\frac{\hbar g}{q_{0}}\right) \xi(t)
\end{eqnarray}
with correlation function
\begin{eqnarray}
    \langle \mathbf{f}_{\mathcal{Q}}(t) \mathbf{f}_{\mathcal{Q}}(t') \rangle = \left(  \frac{\hbar g}{q_{0}} \right)^{2} \cos(\omega(t-t')),
\end{eqnarray}
where the expected value is taken over the distribution of $\mathbf{f}_{\mathcal{Q}}(t)$, which is analogous to Eq.~\eqref{density} but rescaled with ${\hbar g}/{q_{0}}$,
and the dissipative force,
\begin{eqnarray}
    \mathbf{f}_{\rm diss}(t) = 2\frac{\hbar g^2}{q_{0}^{2}} \int_0^t dt' \mathbf{q}(t') \sin(\omega(t-t'))
    \label{dissipation_force}
\end{eqnarray}

\noindent An analogous equation is obtained for $ \mathbf{q}'(t)$. 
In the semiclassical regime, Newton's second law is modified by an additional fluctuation term originating from quantum fluctuations and a dissipation force with memory.

It is interesting to observe the scaling of the stochastic quantum force $ \mathbf{f}_{\mathcal{Q}}(t) $ with Planck's constant. Substituting the zero point motion of the oscillator we find $ \mathbf{f}_{\mathcal{Q}}(t) \propto \sqrt{\hbar} $, while $ \langle \mathbf{f}_{\mathcal{Q}}(t) \mathbf{f}_{\mathcal{Q}}(t') \rangle \propto \hbar $, making the quantum origin of the fluctuations explicit. Note that $ \hbar $ drops out of the dissipation force $ \mathbf{f}_{\rm diss} $. 

For more general quantum states, additional noise contributions and deterministic forces might arise. In that case, the stochastic variable $ \zeta $ will in general be substituted by a sum of uncorrelated stochastic forces $ \zeta(t) \rightarrow \sum_{i} \zeta_{i}  $ satisfying the independence condition $ \langle \zeta_{i} \zeta_{j} \rangle = g^{2} \delta_{ij} A_{i} $. We now turn to the calculation of the noise for quantum states of interest to optomechanical experiments, notably squeezed-coherent and squeezed-thermal states. 


\subsection{Squeezed-coherent states}

We define squeezed-coherent states as \cite{davidovich1996sub},
\begin{eqnarray}
   \ket{\Psi} =  S(z)\ket{\alpha}
    \label{general_squeezed}
\end{eqnarray}
where $ \alpha = \vert \alpha \vert e^{i\theta} $ is the coherent state amplitude and the squeezing operator reads,
\begin{eqnarray}
   S(z) = e^{\frac{1}{2}z^* a^2+\frac{1}{2} z a^{\dagger 2}} 
\end{eqnarray}
with $ z =re^{i\phi} $. We refer to $ r $ as the squeezing parameter and $ \phi $ as the squeezing phase. Using the identities,
\begin{equation}
\begin{aligned}
S^{\dagger}(r, \phi) a S(r, \phi) & = a \cosh r- a^{\dagger} e^{2 i \phi} \sinh r, \\
S^{\dagger}(r, \phi) a^{\dagger} S(r, \phi) & =a^{\dagger} \cosh r- a e^{-2 i \phi} \sinh r.
\end{aligned}
\end{equation}
the influence functional can be brought to the form,
\begin{equation}
  \mathcal{F}_{\mathrm{sq}}[q,q']=  e^{i\Phi_{0}[q,q'] + i\Phi_{0,\mathrm{sq}}[q,q']} F_{\alpha}[q,q'] ,
  \label{squezeed_influence}
\end{equation}
\noindent where,
\begin{eqnarray}
    F_{\alpha}[q,q'] = \bra{\alpha}e^{-a^\dagger f(W)}e^{a f(W)^*} \ket{\alpha}
\end{eqnarray}
and we define,
\begin{equation}
    f(W) = W^*\cosh{r}+We^{2i\phi}\sinh{r},
\end{equation}
In addition to the vacuum phase, squeezed-coherent states acquire a squeezing phase which can be split into a stationary and a non-stationary transient contribution,
\begin{eqnarray}
   i\Phi_{0,\mathrm{sq}}[q,q'] = i\Phi_{0,\mathrm{sq}}^{\rm st}+ i\Phi_{0,\mathrm{sq}}^{\rm n-stat} \label{squezephase}
\end{eqnarray}
with,
\begin{eqnarray}
i\Phi_{0,\mathrm{sq}}^{\rm st} &=& \nonumber - \frac{g^2}{2} (\cosh{2r}-1) \nonumber \\
&\times &\int_0^{\tau}\int_0^{\tau}dtdt' \;J(t)\cos\left({\omega(t-t')} \right) J(t') \nonumber \\
\ 
\end{eqnarray}
and
\begin{eqnarray}
i\Phi_{0,\mathrm{sq}}^{\rm n-st} &=& \nonumber - \frac{g^2}{2} (\sinh{2r}) \nonumber \\
&\times &\int_0^{t_f}\int_0^{t_f}dtdt' \;J(t)\cos\left({\omega(t+t') - 2\phi} \right) J(t') \nonumber \\
\ 
\end{eqnarray}

\noindent The stationary contribution of the influence phase due to squeezing adds to the vacuum phase $ i\Phi_{0}^{\rm fl} $, yielding an enhanced phase proportional to $ \cosh(2r) $. At the same time, we find a non-stationary contribution carrying the information on the squeezing phase $ \phi $. No additional contribution to the dissipation appears due to squeezing. 

We can proceed to perform the Feynman trick for the stationary and non-stationary phase contributions. Introducing auxiliary stochastic variables for each phase in the path integral we arrive at 
\begin{eqnarray}
    \langle \zeta^{\rm st}(t)\zeta^{\rm st}(t')\rangle = g^2\cosh(2r)\cos(\omega(t-t')) 
\end{eqnarray}
for the stationary and
\begin{eqnarray}
    \langle \zeta^{\rm n-stat}(t)\zeta^{\rm n-stat}(t')\rangle =g^2\sinh(2r)\cos(\omega(t+t')-2\phi) \nonumber \\
    \ 
\end{eqnarray}
for the non-stationary terms. Note both stochastic forces are exponentially enhanced in the squeezing parameter, $  \zeta^{\rm st} \propto g\sqrt{\cosh(2r)} $ and $ \zeta^{\rm n-stat} \propto g \sqrt{\sinh(2r)} $. This implies an enhancement in the quantum force due to squeezing, 
 {\begin{eqnarray}
    \mathbf{f}_{\mathcal{Q}}(t) = \mathbf{f}_{\mathcal{Q}}^{\rm st}(t)+\mathbf{f}_{\mathcal{Q}}^{\rm n-st}(t) \nonumber
\end{eqnarray}}
where
\begin{eqnarray}
    \mathbf{f}_{\mathcal{Q}}^{\rm st}(t) &=& \sqrt{\cosh(2r)} \left(\frac{\hbar g}{q_{0}} \right)\xi^{\rm st}(t) \label{stationary_squeezed}\\
    \mathbf{f}_{\mathcal{Q}}^{\rm n-st}(t) &=& \sqrt{\sinh(2r)} \left(\frac{\hbar g}{q_{0}} \right)\xi^{\rm n-st}(t) \label{nonstationary_squeezed}
\end{eqnarray}
with $ \xi^{\rm st}(t) $ and $ \xi^{\rm n-st}(t) $ denote the stationary and non-stationary dimensionless random force variables with correlation functions
\begin{eqnarray}
    \langle \xi^{\rm st}(t) \xi^{\rm st}(t') \rangle &=& \cos(\omega(t-t')) \label{dimensionless_correlators1} \\
    \langle \xi^{\rm n-st}(t) \xi^{\rm n-st}(t') \rangle &=& \cos(\omega(t+t')-2\phi) 
    \label{dimensionless_correlators2}
\end{eqnarray}

Besides the stationary and non-stationary stochastic forces, the squeezed-coherent state also yields a deterministic force arising from the $ F_{\alpha}[q,q'] $ factor in Eq. \eqref{squezeed_influence}. We have,
\begin{equation}
    F_{\alpha}[q,q'] = e^{-\alpha^* f(W)}e^{\alpha f(W)^*} = e^{i\Phi_{\alpha, \mathrm{sq}}},
\end{equation}
where $ i\Phi_{\alpha, \mathrm{sq}} $ denotes the additional influence phase containing the effects of both the coherent and squeezed nature of the state $ \ket{\Psi}$. 
Direct calculation shows that,
\begin{eqnarray}
   i\Phi_{\alpha,\mathrm{sq}}[q,q']=  \frac{i}{\hbar}  \int_{0}^{\tau}dt \left(   \mathbf{q}(t) - \mathbf{q}'(t) \right) \mathbf{f}_{\Psi}(t)
\end{eqnarray}
where,
\begin{eqnarray}
    \mathbf{f}_{\Psi}(t) = - 2|\alpha| \left(\frac{\hbar g}{q_{0}} \right) \left(\sin\left(\omega t-\theta-2\phi\right)\sinh{r}\right.\nonumber\\ \left.+\cos\left(\omega t +\theta\right)\cosh{r}\right).\label{deterministicforce1}
\end{eqnarray}
Note that $  \mathbf{f}_{\Psi}(t) $ represents a deterministic force enhanced by the coherent state amplitude $ \vert \alpha \vert $ and by the exponential squeezing factors $ \sinh r $ and $ \cosh r $. This deterministic force is also quantum mechanical in origin and $  \mathbf{f}_{\Psi}(t) \propto  \sqrt{\hbar} $. 
It is also interesting to note that this force performs work on the mechanical system. In stochastic thermodynamics this work behaves as a random variable and is described by $W[\mathbf{q}(t)]=\int \mathbf{f}_{\Psi}(t) \cdot \dot{\mathbf{q}}(t) dt$ \cite{oliveira2020classical, bo2019functionals, peliti2021stochastic}. A statistical approach can be employed to characterize this work function \cite{paraguassu2022probabilities, paraguassu2022effects, chernyak2006path}.


\subsection{Squeezed-thermal states}

We define squeezed-thermal states as
\begin{eqnarray}
 \rho_{\mathrm{sq,th}} = S(z) \rho_{\mathrm{th}} S^{\dagger}(z)
\end{eqnarray}
where 
\begin{eqnarray}
    \rho_{\mathrm{th}} = \left(1-e^{-\beta\hbar\omega}\right)\sum_n e^{\beta\hbar\omega n}\ket{n}\bra{n},
\end{eqnarray}
and $\beta$ is the inverse temperature associated to the quantum oscillator.
The influence functional assumes the same form as in Eq. \eqref{squezeed_influence},
\begin{eqnarray}
    \mathcal{F}_{\rm sq, th}[q,q'] = e^{i\Phi_{0} + i\Phi_{0,\mathrm{sq}}  } F_{\rm sq, th}[q,q'] 
    \label{influence_thermal}
\end{eqnarray}
where,
\begin{eqnarray}
   F_{\rm sq, th}[q,q'] &=& \left(1-e^{-\beta\hbar\omega}\right) \nonumber \\
   &\times& \sum_n e^{\beta\hbar\omega n}\bra{n}e^{-a^\dagger f(W)}e^{a f(W)^*}\ket{n}
\end{eqnarray}
This sum can be solved analytically, as showed in \cite{parikh2021signatures}, and the total influence functional \eqref{influence_thermal} is expressed in terms of a phase,  {$ \mathcal{F}_{\rm sq, th}[q,q'] = e^{i\Phi_{\rm sq, th}} $}. The final result again contains stationary and non-stationary terms,
\begin{eqnarray}
i\Phi_{\mathrm{sq, th}}[q,q'] = i\Phi_{\mathrm{sq, th}}^{\rm st}+ i\Phi_{\mathrm{sq, th}}^{\rm n-stat} 
\end{eqnarray}
with 
\begin{eqnarray}
i\Phi_{\mathrm{sq,th}}^{\rm st} &=& \nonumber  -\frac{g^2}{2}\cosh({2r})\coth\left(\frac{\beta \hbar \omega}{2}\right) \nonumber \\
&\times &\int_0^{t_f}\int_0^{t_f}dtdt' \;J(t)\cos\left({\omega(t-t')} \right) J(t') \nonumber \\
\ 
\end{eqnarray}
\
\begin{eqnarray}
i\Phi_{\mathrm{sq,th}}^{\rm n-st} &=& \nonumber - \frac{g^2}{2}\sinh({2r})\coth\left(\frac{\beta \hbar \omega}{2}\right) \nonumber \\
&\times &\int_0^{t_f}\int_0^{t_f}dtdt' \;J(t)\cos\left({\omega(t+t') - 2\phi} \right) J(t') \nonumber \\
\ 
\end{eqnarray}

\noindent Once again, we apply the Feynman trick to obtain the stationary and non-stationary stochastic forces, now appearing with enhancement factors due to to the squeezing and thermal nature of the state,
\begin{eqnarray}
    \mathbf{f}_{\mathcal{Q}}^{\rm st}(t) &=& \left[ \cosh(2r) \coth\left(\frac{\beta \hbar \omega}{2}\right) \right]^{1/2} \left(\frac{\hbar g}{q_{0}} \right)\xi^{\rm st}(t) \nonumber \\
    \ \\
    \mathbf{f}_{\mathcal{Q}}^{\rm n-st}(t) &=& \left[ \sinh(2r) \coth\left(\frac{\beta \hbar \omega}{2}\right) \right]^{1/2} \left(\frac{\hbar g}{q_{0}} \right)\xi^{\rm n-st}(t) \nonumber \\
    \ 
\end{eqnarray}
where  $ \xi^{\rm st}(t) $ and $ \xi^{\rm n-st}(t) $ also satisfy \eqref{dimensionless_correlators1} and \eqref{dimensionless_correlators2}.
No additional contribution to the dissipation and deterministic force appears. This is expected, since the mean value of the field amplitude quadrature is zero for squeezed-thermal states.

    

\subsection{Sum over modes}

So far, we have dealt only with the noise contribution arising from a single mode of the optical field. In many applications, we are interested in considering multimode systems. The Feynman-Vernon influence functional then acquires a contribution from each mode \cite{caldeira1983path}. 
Consider for example a multimode cavity. In a discrete $ N $-mode approximation, the Hamiltonian is schematically written as
\begin{eqnarray}
    H_{c}/\hbar = \sum_{k}^{N} \omega_{k} a^{\dagger}_{k} a_{k}
\end{eqnarray}
where $ k = \omega / c $, and the quantum state of the cavity reads,
\begin{eqnarray}
    \vert \Psi \rangle = \bigotimes_{k}^{N} \vert \psi_{\omega_{k}} \rangle
\end{eqnarray}
The Feynman-Vernon influence functional becomes, 
\begin{equation}
    \mathcal{F}[q,q'] = \prod^N_{k} \mathcal{F}_{\omega_{k}}[q,q'].
\end{equation}
In the continuum limit, we must take into account the cavity density of states $ \mathcal{N}(\omega) $ \cite{caldeira1983path}. The total influence phase is then,
\begin{equation}
    \Phi[q,q'] = \sum^{N}_{k} \Phi_{\omega_{k}}[q,q'] \xrightarrow{N \rightarrow \infty} \int_{0}^{\infty} d\omega \mathcal{N}(\omega) \Phi_{\omega}[q,q'] 
    \label{total_influence_phase}
\end{equation}

\noindent For an optical cavity, the density of modes is \cite{fox2006quantum},
\begin{equation}
    \mathcal{N}(\omega) = \frac{1}{\pi}\frac{\gamma}{(\omega-\omega_c)^2+\gamma^2},\label{density}
\end{equation}
where $ \omega_{c} $ is the cavity central frequency and $\gamma$ the cavity damping rate.

Note that the interaction strength of the mechanical oscillator with a given cavity mode $ k $ is frequency dependent $ g = g(\omega) $, which must be taken into account in the integration in the r.h.s. of Eq. \eqref{total_influence_phase}. 
The dissipation force \eqref{dissipation_force} becomes
\begin{eqnarray}
    \mathbf{f}_{\rm diss} = \frac{2\hbar}{q_{0}^{2}} \int_{0}^{\infty} d\omega \mathcal{N}(\omega) g^{2}(\omega) \int_{0}^{t} dt' \mathbf{q}(t') \sin(\omega(t-t')) \nonumber \\
    \ \label{dissipation_cavity}
\end{eqnarray}

\noindent In the case of squeezed-coherent states, the deterministic force reads,
\begin{eqnarray}
    \mathbf{f}_{\Psi}(t) =  -2|\alpha|\frac{\hbar}{q_{0}}\int_0^\infty  d\omega \mathcal{N}(\omega)g(\omega)\left(\cos\left(\omega t +\theta\right)\cosh{r}\right.\nonumber\\\left.+\sin\left(\omega t-\theta-2\phi\right)\sinh{r}\right) \nonumber \\
    \ 
\end{eqnarray}

\noindent Lastly, the stationary and non-stationary noise correlators are, 
\begin{eqnarray}
    \langle \mathbf{f}_{\mathcal{Q}}^{\rm st}(t)\mathbf{f}_{\mathcal{Q}}^{\rm st}(t')\rangle =  \left(\frac{\hbar}{q_{0}}\right)^{2} \int_0^\infty d\omega \mathcal{N}(\omega) g^{2}(\omega) \nonumber \\ 
    \times \cos\left({\omega(t-t')}\right) \label{stat_noise} \\
    \langle \mathbf{f}_{\mathcal{Q}}^{\rm n-st}(t) \mathbf{f}_{\mathcal{Q}}^{\rm n-st}(t')\rangle =  \left(\frac{\hbar}{q_{0}}\right)^{2} \int_0^\infty d\omega \mathcal{N}(\omega) g^{2}(\omega) \nonumber \\ \times \cos\left({\omega(t+t')} - 2\phi \right) \label{non_stat_noise}
\end{eqnarray}
where, in the case of squeezed and thermal states the appropriate enhancement factors must be included. Note that this sum over modes approach can be complemented by an open quantum system model of a cavity interacting with free electromagnetic modes within the formalism of path integrals \cite{neto1990quantum} in an analogous fashion to the quantum Langevin equations \cite{gardiner1985input}.


\section{Semiclassical particle as probe of quantum light}\label{trapedparticle}



We are now in the position to apply the Feynman-Vernon theory to a semiclassical levitated nanoparticle in an optical cavity interacting with a quantum light reservoir via coherent scattering. 


\subsection{Optomechanical parameters}

The coherent scattering coupling rate is given by \cite{delic2019cavity}
\begin{eqnarray}
    \hbar g   =   \alpha  \mathcal{E}_{0}   \mathcal{E}_{c}      k   q_{0} \sin(k \mathbf{r}_{0}) \sin \theta
    \label{g_def}
\end{eqnarray}
where $ \alpha = 3\epsilon_{0}V\left( \frac{\epsilon_{r} - 1}{\epsilon_{r} + 2}  \right) $ is the polarizability of a dielectric particle of volume $ V $, refractive index $ n $ and relative permittivity $ \epsilon_{r} \approx n^{2} $, $ \mathcal{E}_{0} $ is the tweezer field,  {$k$ is the wave number}, $ \mathbf{r}_{0} $ is the mean particle position within the cavity, $ \theta$ is the angle between the tweezer polarization and the cavity axis and $ \mathcal{E}_{c} $ is the cavity electric field strength, given by
\begin{eqnarray}
    \mathcal{E}_{c} = \sqrt{\frac{\hbar \omega}{2\epsilon_{0}V_{c}}} \label{single_photon_field}
\end{eqnarray}
where  {$\omega$ is the frequency and} $ V_{c} $ is the cavity mode volume. We will assume the particle is placed at a cavity node and the tweezer polarization is orthogonal to the cavity axis such that $  \sin(k\mathbf{r}_{0})  = \sin \theta = 1 $; placing the particle at a node only mildly affects the cavity finesse \cite{delic2019cavity}, and hence we will neglect any additional losses introduced by the particle.
We note also that the coherent scattering interaction strength can be tuned and switched on and off by placing the particle at different positions within the cavity or by controlling the polarization of the tweezer field \cite{brandao2021coherent}.

For a confocal cavity of length $ L $ the mode volume in Eq. \eqref{single_photon_field} is given by $ V_{c} = \mathrm{w}_{c}^{2} \pi L / 4 $, where the cavity waist is $ \mathrm{w}_{c} = \sqrt{cL/\omega}  $. In terms of the mode frequency $ \omega $ the optomechanical coupling rate then reads
\begin{eqnarray}
    g \equiv g(\omega) = a \omega^{2} 
    \label{coupling_scaling}
\end{eqnarray}
where 
\begin{eqnarray}
    a = \frac{\alpha \mathcal{E}_{0}}{\sqrt{\pi \epsilon_{0} c^{3} L^{2} m\omega_{m}}}  
\end{eqnarray}
Note that $ g $ is independent of $ \hbar $ as it drops out of Eq. \eqref{g_def}. 

We will find it convenient to define the optomechanical coupling rate at the cavity central frequency,
\begin{eqnarray}
    g_{c} \equiv g(\omega_{c})
\end{eqnarray}
as well as the dimensionless frequency ratios,
\begin{eqnarray}
    \varepsilon &\equiv& g_{c} / \omega_{m} \\
    \nu &\equiv& \gamma / \omega_{c}
\end{eqnarray}
and the characteristic force associated to the central optomechanical coupling,
\begin{eqnarray}
    \mathbf{f}_{0} \equiv \hbar g_{c} / q_{0}
    \label{characteristic_force}
\end{eqnarray}

Table \ref{parameter_table} shows a list of the parameters we will assume for the cavity optomechanical system, similar to the ones described in \cite{delic2019cavity}. Note the dimensionless parameters $ \varepsilon, \nu  \ll 1 $.

\begin{table}[]
\begin{tabular}{cccc}
\hline \hline
Parameter  &   Symbol    & Units & Value \\ \hline 
Cavity length &  $ L $ &   \SI{}{cm}    &  3.0 \\
Cavity central frequency &  $ \omega_{c} $ &   \SI{}{PHz}    &  1.22 \\
Cavity linewidth &  $ \gamma $ &   \SI{}{kHz}    &  $2\pi \times 193$ \\
\\
Particle mass & $ m $ &    \SI{}{fg} & $2.8$ \\
Mechanical frequency & $ \omega_{m} $ &    \SI{}{kHz} & $2\pi \times 190$ \\
Zero point fluctuation & $ q_{0} $ &    \SI{}{m} & $ 3.6\times 10^{-12} $ \\
Damping rate at & $ \gamma_{m} $  &    \SI{}{Hz} & $5 \times 10^{-6} $ \\
Coupling rate at $ \omega_{c} $ & $ g_{c} $ &    \SI{}{kHz} & $2\pi \times 18 $ \\
\\
Coupling-to-mech. freq. ratio  & $ \varepsilon $ &  - &  $0.1$  \\
Cavity linewidth-to-freq. ratio  & $ \nu $ &  - &  $10^{-9}$  \\
    \hline \hline 
\end{tabular}
\caption{\label{parameter_table} Coherent scattering optomechanical parameters. Values adapted from \cite{delic2019cavity}.}
\end{table}



\subsection{Vacuum fluctuations and dissipation}




We can calculate the stochastic force $ \mathbf{f}_{\mathcal{Q}}(t) $ associated to the vacuum fluctuations in the optical cavity using Eq. \eqref{stat_noise}. In this case we only have the stationary process with noise kernel,
\begin{eqnarray}
    \langle \mathbf{f}_{\mathcal{Q}}(t)\mathbf{f}_{\mathcal{Q}}(t')\rangle = \frac{\gamma}{\pi}\left(\frac{\hbar a}{q_{0}}\right)^{2} \int_{0}^{\infty} d\omega \frac{\omega^4\cos(\omega(t-t'))}{(\omega-\omega_c)^2+\gamma^2}
    \label{vacuum_kernel}
\end{eqnarray}
where we omit the stationary `st' superscript for simplicity. The frequency integral \eqref{vacuum_kernel} can be solved in terms of distributions -- we refer to Appendix \ref{integralsapp} for the details. We find that the vacuum state introduces three independent stationary stochastic force components,
\begin{eqnarray}
    \mathbf{f}_{\mathcal{Q}}(t) = \mathbf{f}_{0} \sum_{i} \xi_{i}(t)
    \label{quantum_fluct_force}
\end{eqnarray}
with the correlation functions
\begin{eqnarray}
    \langle \xi_{i}(t) \xi_{j}(t') \rangle = \delta_{ij} A_{i}(\tau)
    \label{quantum_correlators}
\end{eqnarray}
where, $ \tau \equiv t - t' $ and
\begin{eqnarray}
    A_{1}(\tau) = e^{-\gamma |\tau|} \left(  (1 - 6\nu^{2} + \nu^{4}) \cos(\omega_{c}\tau) \right. \nonumber \\ 
    \left. - 4 (\nu - \nu^{3}) \sin(\omega_{c}|\tau|)   \right)
    \label{quantum_correlators1}
\end{eqnarray}
\begin{eqnarray}
    A_{2}(\tau) &=& \left( 3\nu - \nu^{3}   \right) \frac{\delta(\tau)}{\omega_{c}} \label{quantum_correlators2} \\
     A_{3}(\tau) &=& -\nu \frac{\delta''(\tau)}{\omega_{c}^{3}} \label{quantum_correlators3}
\end{eqnarray}

\noindent The correlator in Eq. \eqref{quantum_correlators1} resembles active noise with a relative phase shift between the sine and cosine components \cite{dabelow2019irreversibility, caprini2021inertial}. Colored noise can be generally expressed as an expansion in terms of derivatives of the delta distribution \cite{tome2015stochastic}, hence the second and third terms represent additional colored noise. Moreover, the second derivative delta noise has appeared previously in the context of stochastic gravity \cite{Beilok1995}.

The leading contribution to the stochastic force correlator comes from the zeroth order term in $ \nu $. We approximate the stochastic force correlator as
\begin{align}
    \langle \mathbf{f}_{\mathcal{Q}}(t)\mathbf{f}_{\mathcal{Q}}(t') \rangle \approx \mathbf{f}_{0}^{2} \ e^{-\gamma \vert \tau \vert} \cos(\omega_{c} \tau) 
    \label{approximate_correlator}
\end{align}
Observe the characteristic scale of the stochastic force is given by $ \mathbf{f}_{0} $.

We now turn to the dissipation. Substituting Eqs. \eqref{density} and \eqref{coupling_scaling} into \eqref{dissipation_cavity} we arrive at
\begin{eqnarray}
    \mathbf{f}_{\rm diss} = \frac{2}{\pi}\frac{\hbar \gamma}{q_{0}} a^{2} \int_{0}^{t} dt' q(t') \int_{0}^{\infty} d\omega \frac{\omega^4\sin(\omega \tau)}{(\omega-\omega_c)^2+\gamma^2}
\end{eqnarray}
Once again, the frequency integral can be evaluated in terms of distributions, see Appendix \ref{integralsapp} for details. We find the dissipation force,
\begin{eqnarray}
    \mathbf{f}_{\rm diss} = \mathbf{f}_{0} \left[ 12 \nu \left( \frac{g_{c}}{\omega_{c}}  \right) \frac{\dot{q}(t)}{\omega_{c}} + \varepsilon u(t) \right]
    \label{final_dissipation}
\end{eqnarray}
where
\begin{align}
    u(t) = 2\int_{0}^{t} d(\omega_{m}t') e^{-\gamma |\tau|} \left(  (1 - 6\nu^{2} + \nu^{4}) \sin(\omega_{c}\tau) \right. \nonumber \\
    \left. + 4(\nu - \nu^{3}) \mathrm{sgn}(\tau) \cos(\omega_{c}\tau)    \right) q(t')
\end{align}
We see the total dissipation force $ \mathbf{f}_{\rm diss} $ consists of a standard velocity-dependent term plus a modification of the mechanical spring constant with a memory kernel.

The dissipation is small when compared to the fluctuation force. From Table \ref{parameter_table}, $ g_{c} \approx 0.1 \gamma $, while $ \omega_{m} \approx \gamma $. Moreover, $ \dot{q}(t) \approx \omega_{m} q(t) $. Hence, the first term in Eq. \eqref{final_dissipation} is $ \mathcal{O}(\nu^{3}) $, while the second term is at most of order $ \mathcal{O}(\varepsilon) $ in units of $ \mathbf{f}_{0} $. We will henceforth neglect dissipation effects arising from the influence functional.

\subsection{Particle dynamics}

We arrive at the equation of motion for the mechanical oscillator,
\begin{align}
    m\ddot{\mathbf{q}} + \Gamma_{m}\dot{\mathbf{q}} + m\omega_{m}^{2} \mathbf{q} = \mathbf{f}_{\mathcal{Q}}(t) + \mathbf{f}_{\mathrm{diss}} + \eta(t)
    \label{cavity_langevin}
\end{align}
where we have added phenomenologically a dissipation with damping coefficient $ \Gamma_{m} $ and the thermal white noise $ \eta(t) $ with 
\begin{eqnarray}
    \langle \eta(t) \eta(t') \rangle = 2\Gamma_{m}k_{B}T_{\rm bath} \delta(\tau)
\end{eqnarray}
where $ T_{\rm bath} $ is the temperature of the particle's environment. Note the stochastic quantum force $ \mathbf{f}_{\mathcal{Q}}(t) $ might contain stationary and non-stationary components and is accompanied by a state-dependent correlation function. For example, for the vacuum state, the correlation function of $ \mathbf{f}_{\mathcal{Q}}(t) $ is given in Eq. \eqref{approximate_correlator}.

Considering only the zeroth order terms in the dimensionless parameters $ \nu $ and $\varepsilon$, the equation of motion simplifies to
\begin{align}
    m\ddot{\mathbf{q}} + \Gamma_{m}\dot{\mathbf{q}} + m\omega_{m}^{2} \mathbf{q} \approx \mathbf{f}_{\mathcal{Q}}(t) + \eta(t)
    \label{cavity_langevin}
\end{align}
We have the formal solution
\begin{align}
    \mathbf{q}(t) = \frac{e^{-\gamma_{m}t}}{m\Omega_{m}} \int_{0}^{t} ds e^{\gamma_{m}s}\sin\left( \Omega_{m}(s-t)  \right) \left(  \mathbf{f}_{\mathcal{Q}}(s) + \eta(s)   \right)
    \label{formal_solution}
\end{align}
where,
\begin{eqnarray}
    \gamma_{m} &\equiv& \Gamma_{m} / 2m \\
    \Omega_{m} &\equiv& \sqrt{\omega_{m}^{2} - \gamma_{m}^{2}}
\end{eqnarray}
and, for simplicity, we assume initial conditions $  \mathbf{q}(0) = \dot{\mathbf{q}}(0) = 0 $. For a particle with a radius of $ 70 $ nm at pressures around $ 10^{-9} $ mbar, $ \gamma_{m} \approx 5 \times 10^{-6} $ Hz so we have $ \Omega_{m} \approx \omega_{m} $ \cite{magrini2021real}.

Let $ \sigma_{\mathbf{q}} \equiv \sqrt{\langle \mathbf{q}^{2}(t) \rangle} $ be the particle's position root-mean-square (rms).
We are interested in calculating $ \sigma_{\mathbf{q}} $ while the particle interacts with a quantum state populating the cavity and an external thermal reservoir. 
The position rms $ \sigma_{\mathbf{q}}^{2} $ then has two independent contributions, one from the quantum force $ \mathbf{f}_{\mathcal{Q}}(t) $ and the other from the uncorrelated thermal noise $ \eta(t) $. We define the \textit{excess} quantum-induced fluctuations as 
\begin{eqnarray}
    \Delta \sigma_{\mathbf{q}}^{2} = \sigma_{\mathbf{q}}^{2} - \sigma_{0}^{2} \label{excessrms}
\end{eqnarray}
where $ \sigma_{0}^{2} $  denotes the non-quantum contribution to $ \sigma_{\mathbf{q}}^{2} $ arising from the thermal fluctuations $ \eta(t) $ and
\begin{eqnarray}
   \Delta \sigma_{\mathbf{q}}^{2} = \frac{e^{-2\gamma_{m}t}}{m^{2}\Omega_{m}^{2}}  \int_{0}^{t} \int_{0}^{t} ds ds' e^{\gamma_{m}(s+s')}  \nonumber \\
    \times \sin\left( \Omega_{m}(s-t)  \right) \sin\left( \Omega_{m}(s'-t)  \right) \langle \mathbf{f}_{\mathcal{Q}}(s)\mathbf{f}_{\mathcal{Q}}(s') \rangle
    \label{excess_variance_mechanical}
\end{eqnarray}
The solution to this integral is exact but cumbersome. In essence, $\Delta \sigma_{\mathbf{q}} $ reaches a steady state with characteristic value is given by
\begin{align}
    \Delta \sigma_{\mathbf{q}} \approx  \frac{\mathbf{f}_{0}}{m\sqrt{\Omega_{m}^{3}\omega_{c}}}
    \label{standard_vacuum}
\end{align}
As we will see in Sec. \ref{effects}, this quantum contribution to the particle's position rms is negligible for current experiments with levitated nanoparticles.

\subsection{Squeezed-coherent states}

We now consider squeezed-coherent states. Due to their displacement in phase space, these states introduce a deterministic force on the particle given by,
\begin{align}
    \mathbf{f}_{\psi} = - \frac{2\vert \alpha \vert }{\pi} \frac{\hbar \gamma}{q_{0}} a \int_{0}^{\infty} d\omega \frac{\omega^2}{(\omega-\omega_c)^2+\gamma^2} \nonumber \\
    \times \left(\cos\left(\omega t +\theta\right)\cosh{r} 
 + \sin\left(\omega t-\theta-2\phi\right)\sinh{r} \right)
\end{align}
Again, this integral can be solved following the steps in Appendix \ref{integralsapp}.
For simplicity, consider a real coherent state amplitude ($\theta = 0$).  {To zeroth order in $\nu$} the deterministic force reads
\begin{eqnarray}
    \mathbf{f}_{\psi} =  \mathbf{f}_{0} \vert \alpha \vert \  e^{-\gamma t} \left[  
    2 \cosh r  \cos(\omega_{c}t) \right. \nonumber \\
     \left. + \sinh r \left(  \sin 2\phi  \sin(\omega_{c}t) - 2\cos 2\phi  \cos(\omega_{c}t) \right) \right]
\end{eqnarray}
This is a fast oscillating force, which time averages to zero. However, note that it exhibits an exponential enhancement due to squeezing and depending on the squeezing phase this force can be exponentially enhanced or suppressed. The same enhancement/suppression effects as a function of squeezing phase have been shown to appear in the context of a quantized gravitational wave in a squeezed-coherent state interacting with an optical cavity \cite{guerreiro2022quantum}.

The stationary noise associated to squeezing assumes the same form as the vacuum, but is enhanced exponentially in the squeezing parameter,
\begin{eqnarray}
    \mathbf{f}_{\mathcal{Q}}^{\mathrm{st}}(t) = \sqrt{\cosh(2r)} \  \mathbf{f}_{0} \ \sum_{i} \xi_{i}(t)
    \label{quantum_fluct_force}
\end{eqnarray}
where the dimensionless stochastic variables $ \xi_{i}(t) $ satisfy Eqs. \eqref{quantum_correlators}-\eqref{quantum_correlators3}. The noise correlation function to zeroth order in $ \nu $ is approximated by
\begin{align}
    \langle \mathbf{f}_{\mathcal{Q}}^{\mathrm{st}}(t)\mathbf{f}_{\mathcal{Q}}^{\mathrm{st}}(t') \rangle \approx \cosh(2r) \ \mathbf{f}_{0}^{2} \ e^{-\gamma \vert \tau \vert} \cos(\omega_{c} \tau) 
    \label{approximate_squeezed_correlator}
\end{align}

Squeezed-coherent states also exhibit the non-stationary noise defined in Eq. \eqref{non_stat_noise}. To zeroth order in $\nu$, 
\begin{align}
    \langle \mathbf{f}_{\mathcal{Q}}^{\rm n-st}(t)\mathbf{f}_{\mathcal{Q}}^{\rm n-st}(t') \rangle \approx \sinh(2r)\ \mathbf{f}_{0}^{2} \ e^{-\gamma (t+t')} \nonumber \\\times \left(\frac{\cos(2\phi)}{2}\cos(\omega_{c} (t+t'))+\sin(2\phi)\sin(\omega_{c} (t+t'))\right)
\end{align}
This non-stationary noise is also exponentially enhanced, despite decaying with a characteristic time given by $\gamma^{-1}$. Note, however, that $ \mathbf{f}_{\mathcal{Q}}^{\mathrm{n-st}}(t) $ depends on the squeezing angle $ \phi $, and is maximized for $ \phi = \pi/2 $. In that case, for short times $ \mathbf{f}_{\mathcal{Q}}^{\mathrm{n-st}}(t) \approx \mathbf{f}_{\mathcal{Q}}^{\mathrm{st}}(t) $ and the effects of the non-stationary and stationary noises becomes comparable. 

Finally, no additional contribution to the dissipation arises from squeezing. We conclude that for long times, the squeezed-coherent state contributes to the particle's rms as
\begin{eqnarray}
    \Delta \sigma_{\mathbf{q}} \approx \sqrt{\cosh(2r)}  \frac{\mathbf{f}_{0}}{m\sqrt{\Omega_{m}^{3}\omega_{c}}}   
    \label{standard_squeezed}
\end{eqnarray} 

\subsection{Squeezed-thermal states}

Squeezed-thermal states also present stationary and non-stationary noise defined by, 
    \begin{align}
    \langle \mathbf{f}_{\mathcal{Q}}^{\mathrm{st}}(t)\mathbf{f}_{\mathcal{Q}}^{\mathrm{st}}(t')\rangle = \cosh(2r)\frac{\gamma}{\pi}\left(\frac{\hbar a}{q_{0}}\right)^{2} \nonumber \\
    \times \int_0^\infty d\omega \frac{\omega^4\coth\left(\frac{\beta \hbar \omega}{2}\right)\cos\left({\omega \tau }\right)}{(\omega-\omega_c)^2+\gamma^2},\label{statthermal2}\\
    \langle \mathbf{f}_{\mathcal{Q}}^{\mathrm{n-st}}(t)\mathbf{f}_{\mathcal{Q}}^{\mathrm{n-st}}(t')\rangle = \sinh(2r)\frac{\gamma}{\pi}\left(\frac{\hbar a}{q_{0}}\right)^{2}  \nonumber \\
    \times \int_0^\infty d\omega \frac{\omega^4\coth\left(\frac{\beta \hbar \omega}{2}\right)\cos\left({\omega(t+t')-2\phi}\right)}{(\omega-\omega_c)^2+\gamma^2}\label{nonstatthermal}
\end{align}
 {where $\beta = 1/k_BT$}.
In the high temperature limit $\coth({\beta \hbar \omega/2})\rightarrow 2/(\beta \hbar\omega)$ and the integrals can be solved analytically, following similar steps given in the Appendix \ref{integralsapp}. We find for the stationary noise,
\begin{eqnarray}
    \mathbf{f}_{\mathcal{Q}}^{\mathrm{st}}(t) = \left[\cosh(2r) \left(\frac{k_{B}T}{\hbar\omega_{c}}\right)\right]^{1/2}  \mathbf{f}_{0}   \sum_{i} \xi_{i}(t)
    \label{quantum_fluct_force}
\end{eqnarray}
where $ \xi_{i = 1,2} $ are independent random variables with correlators,
\begin{eqnarray}
    A_{1}(\tau) = 2e^{-\gamma |\tau|} \left(  (1 - 3\nu^{2} ) \cos(\omega_{c}\tau) \right. \nonumber \\ 
    \left. -  (\nu - \nu^{3}) \sin(\omega_{c}|\tau|)   \right)
\end{eqnarray}
\begin{eqnarray}
    A_{2}(\tau) = 8 \nu \frac{\delta(\tau)}{\omega_{c}}
\end{eqnarray}
Once again, considering only the zeroth order terms in $ \nu $ we can approximate the stationary noise correlator as
\begin{align}
    \langle \mathbf{f}_{\mathcal{Q}}^{\mathrm{st}}(t)\mathbf{f}_{\mathcal{Q}}^{\mathrm{st}}(t') \rangle \approx 2\cosh(2r)\left(\frac{k_{B}T}{\hbar\omega_{c}}\right)  \mathbf{f}_{0}^{2} \ e^{-\gamma \vert \tau \vert} \cos(\omega_{c} \tau) 
    \label{approximate_squeezed_correlator}
\end{align}
We see that besides the exponential squeezing enhancement the thermal state further increases the noise in proportion to its temperature. 

Moving on to the non-stationary noise, we find to zeroth order in $\nu$,
\begin{align}
    \langle \mathbf{f}_{\mathcal{Q}}^{\mathrm{n-st}}(t)\mathbf{f}_{\mathcal{Q}}^{\mathrm{n-st}}(t') \rangle \approx \sinh(2r) \left(\frac{k_{B}T}{\hbar\omega_{c}}\right) \mathbf{f}_{0}^{2} \ e^{-\gamma(t+t')} 
    \nonumber \\
    \times \left[  2\cos(2\phi) \cos(\omega_{c}(t+t')) - \sin(2\phi) \sin(\omega_{c}(t+t'))   \right]
\end{align}

For the squeezed-thermal state the particle's position rms is modified to
\begin{eqnarray}
    \Delta \sigma_{\mathbf{q}} \approx \left[ 2\cosh(2r)\left(\frac{k_{B}T}{\hbar\omega_{c}}\right) \right]^{1/2} \frac{\mathbf{f}_{0}}{m\sqrt{\Omega_{m}^{3}\omega_{c}}}   
    \label{standard_thermal}
\end{eqnarray}
presenting an enhancement both due to squeezing and temperature of the optical reservoir.

In the absence of squeezing we can show that the fluctuation-dissipation relation holds \cite{vankampen1989langevin}. Define
\begin{eqnarray}
    T(t,t') = \frac{2}{\pi} \frac{\gamma}{\hbar \omega_{c}^{4}} \ \mathbf{f}_{0}^{2} \int_{0}^{\infty} d\omega \frac{\omega^3\cos(\omega\tau)}{(\omega-\omega_c)^2+\gamma^2} \label{diss_kernel}
\end{eqnarray}
and observe that Eq. \eqref{dissipation_cavity} can be written as,
\begin{eqnarray}
    \mathbf{f}_{\mathrm{diss}} &=&  \int_{0}^{t} \dot{T}(t,t')\mathbf{q}(t')dt' \nonumber \\
    &=& - \int_{0}^{t} T(t,t') \dot{\mathbf{q}}(t') dt'
\end{eqnarray}
where we have integrated by parts and assumed vanishing boundary terms. Moreover, from Eqs. \eqref{statthermal2} and \eqref{diss_kernel} we have
\begin{eqnarray}
    \langle \mathbf{f}_{\mathcal{Q}}^{\mathrm{st}}(t)\mathbf{f}_{\mathcal{Q}}^{\mathrm{st}}(t') \rangle  = \frac{1}{\beta} T(t,t')
\end{eqnarray}
Hence, for a high occupation thermal state the dissipation and fluctuation kernels are related as usual.

\subsection{Quantitative estimations}\label{effects}

We can now estimate how large is the quantum contribution to the particle's position rms due to interaction with an optical reservoir in the vacuum, squeezed-coherent and squeezed-thermal states given the optomechanical parameters in Table \ref{parameter_table}.

The characteristic value of the quantum stochastic force is 
\begin{eqnarray}
\mathbf{f}_{0} \approx 10^{-18} \ \mathrm{N}
\end{eqnarray}
which is about 100 times stronger than the gravitational force between two Planck masses at a 1 mm separation distance \cite{westphal2021measurement}. For the vacuum, the parameters in Table \ref{parameter_table} give
\begin{align}
    (\Delta \sigma_{\mathbf{q}})_{\mathrm{vac}} \approx  2\times10^{-6} q_{0} \ ,
    \label{vacuum_value}
\end{align}
well below the zero point fluctuations of the mechanical ground state and hence negligible. 

The exponential enhancement provided by squeezing can be used to improve \eqref{vacuum_value}. For example, for a squeezing parameter of $r = 14 $, corresponding to $\approx 60$ dB of quadrature squeezing, we find
\begin{align}
    (\Delta \sigma_{\mathbf{q}})_{\mathrm{sq}} \approx 2q_{0}
    \label{squeezed_value}
\end{align}
This is arguably an enormous amount of squeezing, to be compared to state-of-the-art squeezing sources operating producing states around 10 dB \cite{vahlbruch2008observation}. 

We can relax the amount of squeezing if we populate the cavity with a squeezed-thermal state. For instance, a state with $ k_{B}T/\hbar \omega_{c} = 10^{5} $ and 12 dB squeezing produces $ (\Delta \sigma_{\mathbf{q}})_{\mathrm{sq, th}} \approx 2q_{0} $. 

\section{Semiclassical Light as probe of quantum particle \label{sec:light-particle}}

So far, we have focused on the semiclassical equations of motion for the mechanical oscillator interacting with the optical cavity reservoir in a quantum mechanical state. However, the linear optomechanical Hamiltonian  {could be} symmetric in the mechanical and optical degrees of freedom. With the appropriate modifications, we can use the methods developed so far to describe a semiclassical optical field interacting with a quantum mechanical oscillator. 

\subsection{Optical equations of motion}

The derivation of the optical semiclassical equations of motion is very analogous to that of the mechanical case and with the proper substitution of constants we can fast-forward to the optical version of the results in Sec. \ref{sec:eq_motion}.
For simplicity, we will consider a single cavity mode. The Hamiltonian describing our system is the same as in Eq. \eqref{total_hamiltonian}. Up to a constant shift, the free optical Hamiltonian in Eq. \eqref{optical_hamiltonian} can be written as
\begin{eqnarray}
    H_{c}/\hbar = \frac{\omega}{4}\left( X(t)^{2} + Y(t)^{2}   \right)
    \label{quadratures_hamiltonian}
\end{eqnarray}
where the optical field quadratures are defined as
 {
\begin{eqnarray}
    X(t) &=& \frac{a e^{-i\omega t} + a^{\dagger} e^{i\omega t}}{\sqrt{2}}\\
    Y(t) &=& \frac{i \left( a^{\dagger} e^{i\omega t} - a e^{-i\omega t} \right)}{\sqrt{2}}
\end{eqnarray}}
and $ \dot{X} = \omega Y $. We define the electromagnetic position and momentum quadratures 
\begin{eqnarray}
    Q \equiv \sqrt{\frac{\hbar}{m_{0}\omega}} X \ \  , \  \ P = m_{0} \dot{Q}
\end{eqnarray}
where $ m_{0} $ is a constant with dimension of mass introduced to establish the analogy between an optical mode and a harmonic oscillator \cite{scully1999quantum}. In terms of $ Q, P $ the Hamiltonian \eqref{quadratures_hamiltonian} becomes
\begin{eqnarray}
    H_{c} = \frac{1}{2}\left(  m_{0}\omega^{2}Q^{2} + \frac{P^{2}}{m_{0}}   \right)
\end{eqnarray}
with the corresponding Lagrangian,
\begin{eqnarray}
    L_{c} &=& \frac{m_{0}}{2} \left(   \dot{Q}^{2} - \omega^{2} Q^{2} \right) \nonumber \\
    &=& \frac{\hbar}{4\omega} \left(   \dot{X}^{2} - \omega^{2} X^{2} \right) \label{electromagnetic_lagrangian}
\end{eqnarray}
Note that $m_{0}$ drops out of $ L_{c} $. 

Repeating the steps up to Sec. \ref{sec:eq_motion} we obtain the electromagnetic version of the stochastic propagator in Eq. \eqref{final__mechanical_propagator} after tracing out the mechanical degree of freedom. For the mechanical oscillator initially in the ground state we have
\begin{widetext}
    \begin{align}
        \mathcal{J}(X_t,X'_t|X_0,X'_0)  = \int_{X_0,X'_{0}}^{X_t,X'_t} \mathcal{D}X\mathcal{D}X' \int \mathcal{D}\zeta  P[\zeta(t)]
        \exp\left(\frac{i}{\hbar}\int_{0}^{t_f} dt \left(L_{c} - L_{c}'\right)\right)\nonumber \\
        \exp\left(\frac{i}{\hbar}\int_0^{t_f} dt \;
         \hbar \zeta(t) (X(t)-X'(t))+ \frac{i}{\hbar} \left(\hbar g^{2} \right) \int_0^{t_f}\int_0^{t}dt\, dt'\left(X(t)-X'(t)\right)\left(X(t')+X'(t')\right)\sin\left(\omega_{m}(t-t')\right)\right)
        \label{final_optical_propagator}
    \end{align}
\end{widetext}
where the probability density $ P[\zeta(t)] $ is defined in \eqref{density} and $ L_{c}' $ is the Lagrangian in Eq. \eqref{electromagnetic_lagrangian} in terms of the backward quadrature $ X \rightarrow X'$. Note that we can express $\zeta(t)$ in terms of the dimensionless random variable $\xi(t)$ by using Eq.~(\ref{dimensionnoise}).

From \eqref{final_optical_propagator} we can calculate the semiclassical Eqs. of motion for the cavity field quadrature. As in the mechanical case we neglect the coupling between the forward and backward variables $ X $ and $ X' $. We have,
\begin{eqnarray}
    \ddot{X} + \gamma \dot{X} + \omega^{2} X = \mathbf{F}_{Q}(t) + \mathbf{F}_{\mathrm{diss}}(t)
    \label{optical_eq}
\end{eqnarray}
where $ \mathbf{F}_{Q}(t) $ is the stochastic ``force'' arising from the quantum fluctuations of the mechanical oscillator,
\begin{eqnarray}
    \mathbf{F}_{Q}(t) = 2 \omega g \xi(t)
\end{eqnarray}
with
\begin{eqnarray}
    \langle \xi(t) \xi(t') \rangle = \cos(\omega_{m}\tau)
    \label{dimensionless_mechanical_correlator}
\end{eqnarray}
and $ \mathbf{F}_{\mathrm{diss}}(t) $ is the dissipation term given by
\begin{eqnarray}
    \mathbf{F}_{\mathrm{diss}}(t) = 2\omega g^{2} \int_{0}^{t} dt' X(t') \sin(\omega_{m}(t-t'))
\end{eqnarray}
Note we have introduced a phenomenological dissipative term $ \gamma \dot{X} $ due to the coupling of the cavity mode with the external electromagnetic field \cite{milburn2012quantum}. 

To compare the order-of-magnitude of the terms in Eq. \eqref{optical_eq} we rescale time according to $ t \rightarrow \omega_{m} t $. The equation of motion becomes,
\begin{eqnarray}
    \ddot{X} + \left( \frac{\gamma}{\omega_{m}} \right) \dot{X} +  \left( \frac{\omega}{\omega_{m}} \right)^{2} X = 2 \varepsilon \left( \frac{\omega}{\omega_{m}} \right)  \xi(t) \nonumber \\
    + 2 \varepsilon^{2} \left( \frac{\omega}{\omega_{m}} \right)  \int_{0}^{t} dt' X(t') \sin(t-t')
\end{eqnarray}
where here derivatives and integration are taken with respect to rescaled time. Again we see the fluctuation is of order $ \varepsilon $ while the dissipation arising from the influence functional is of order $ \varepsilon^{2} $, which from now on will be neglected. 

In the mechanical case we have summed over the cavity modes weighted by the Lorentzian density of states of width $\gamma$. 
Similarly, the mechanical mode also couples to external environmental degrees of freedom and a sum over modes procedure is in order. However, the broadening of the mechanical mode given by the mechanical damping rate $ \gamma_{m} $ is comparatively much smaller than that of the cavity. 
We have $ \gamma_{m} / \omega_{m} \approx 10^{-11} \ll \nu $. We will therefore consider the mechanical oscillator as a single mode system. This approximation is valid for times much smaller than the inverse mechanical damping rate, $ t \ll \gamma_{m}^{-1} $.

We arrive at the final form of the semiclassical optical equations for the field quadratures,
\begin{eqnarray}
    \dot{X} &=& \omega Y \\
    \dot{Y} &\approx& - \omega X - \gamma Y + 2g \xi(t) 
\end{eqnarray}
where $ \xi(t) $ satisfies \eqref{dimensionless_mechanical_correlator}.

\subsection{Fluctuations}

Let us estimate the order of magnitude of the fluctuations imprinted on the cavity by a quantum mechanical oscillator in the ground state. For simplicity we neglect the cavity dissipation. In that case the optical equations are formally solved by
\begin{align}
     X(t) = X_0 \cos(\omega t)+ Y_0 \sin(\omega t) \nonumber \\
     + 2g\int_0^t ds \sin(\omega(t-s))\xi(s) ,\\ 
     Y(t) = -X_0\sin(\omega t)+Y_0 \cos(\omega t) \nonumber \\
     +  2g \int_0^t ds \cos(\omega(t-s))\xi(s) .
\end{align}
where the field quadratures have initial conditions $ X(0) = X_{0},  Y(0) = Y_{0} $.
The field quadrature rms reads
\begin{eqnarray}
    \sigma_{X}^{2} = \sigma_{0}^{2} + \Delta\sigma_{X}^{2}
\end{eqnarray}
where 
\begin{eqnarray}
    \sigma_{0}^{2} = \sigma_{X_{0}}^{2} \cos^{2}(\omega t) + \sigma_{Y_{0}}^{2} \sin^{2}(\omega t) \nonumber \\
    + 2\mathrm{Cov}(X_{0},Y_{0})\sin(\omega t)\cos(\omega t)
\end{eqnarray}
and the excess quantum-induced fluctuations are
\begin{eqnarray}
     \Delta\sigma_{X}^{2} = 4 g^{2}\int_{0}^{t}\int_{0}^{t}ds ds' \sin(\omega(t-s)) \nonumber \\
     \sin(\omega(t-s')) \langle \xi(s)\xi(s')\rangle
     \label{optical_variance_integral}
\end{eqnarray}
Note that with the appropriate changes of constants $ \Delta\sigma_{X}^{2} $ assumes the same form as Eq. \eqref{excess_variance_mechanical} with zero damping rate $ \gamma_{m} = 0 $. 

\subsection{Quantitative Estimations}

Since we are interested in the changes in the quadrature rms due to interaction with the mechanical oscillator we need only look into $ \Delta\sigma_{X}^{2} $. We have,
\begin{eqnarray}
        \Delta \sigma_{X}^{2} = \frac{4g^{2} \omega^{2}}{(\omega^{2} - \omega_{m}^{2})^{2}} h(t)
        \label{field_X_quadrature_variance}
    \end{eqnarray}
where
\begin{eqnarray}
    h(t) = 1 + \cos^{2}(\omega t) - 2 \cos(\omega t)\cos(\omega_{m}t) \nonumber \\
    - 2 \kappa \sin(\omega t)\sin(\omega_{m}t) + \kappa^{2} \sin^{2}(\omega_{m}t)
    \label{oscillations1}
\end{eqnarray}
and $ \kappa = \omega_{m} / \omega $.
Similar expressions for $ \sigma_{Y}^{2} $ can be obtained. 
We see that to leading order in $(\omega_{m}/\omega)$, the optical quadrature rms oscillates with a characteristic amplitude  given by $ \Delta \sigma_{X}^{2} \approx (g / \omega)^{2} $, which for the optomechanical parameters in Table \ref{parameter_table} is $ g/\omega \approx 10^{-10} $. This number is to be compared with the standard deviation of the optical coherent state $ \sigma_{X_{0}} = 1/2 $.
Again, ground state fluctuations are too small to have any practical effects in levitated optomechanics. 

For the case of squeezed states, the result in Eq. \eqref{field_X_quadrature_variance} will acquire an exponential enhancement and a non-stationary contribution dependent on the squeezing phase as in the mechanical case. However, to elevate the amplitude of the oscillations in \eqref{field_X_quadrature_variance} to a level comparable to the standard deviation of a coherent state  {will require an  enhancement factor of $\approx e^{23}$}, corresponding to an impractical amount of squeezing. It is interesting to observe that a room-temperature levitated nanoparticle yields an appreciable effect, $ \Delta \sigma_{X} \propto (k_{B}T_{\rm bath} / \hbar \omega_{m}) (g/\omega) \approx 10^{-3}  $ for $ T_{\rm bath} \approx 293 $ K.

As a final observation, note that the rms in Eq. \eqref{field_X_quadrature_variance} has a factor inversely proportional to the square difference of the oscillators' frequencies. For a cavity interacting with a mechanical oscillator the higher optical frequency dominates, leading to a suppression of $ \Delta \sigma_{X} $. If two oscillators with similar frequencies interact via the linear optomechanical Hamiltonian we can expect a resonantly enhanced effect. This motivates our final application of two linearly coupled levitated nanoparticles.

\section{Semiclassical particle as probe of quantum particle \label{sec:particle-particle}}

We turn to two linearly coupled mechanical oscillators. Different coupling mechanisms between levitated nanoparticles have been recently demonstrated, such as the Coulomb interaction \cite{penny2023sympathetic, rieser2022tunable}, optical binding \cite{rieser2022tunable,arita2022all} and cavity-mediated interactions \cite{vijayan2023cavity}.  {It is worth mentioning that nonreciprocal interactions between levitated particles arising from coherent scattering has been reported in \cite{rieser2022tunable, reisenbauer2024non}. Although the methods described in this work could be applied to non-reciprocal interactions, we will consider only reciprocal coupling arising from the electrostatic force.}

From now on modes $ a $ and $ b $ in the Hamiltonian \eqref{total_hamiltonian} are interpreted as two identical harmonically trapped nanoparticles with frequencies $ \omega_{a}, \omega_{b} $. We consider $ \omega_{a} \gtrsim  \omega_{b} $ and trace out mode $ a $, assumed to be the quantum system. For simplicity, we will neglect external decoherence acting on the quantum system, although we note that these effects can also be taken into account using the path integral formalism \cite{neto1990quantum}. We also neglect the mechanical damping $ \gamma_{m} $ and as in Sec. \ref{sec:light-particle}, our conclusions will be valid for times $ t \ll \gamma_{m}^{-1} $.

As an example of coupling mechanism we consider the Coulomb interaction between charged particles. For small displacements with respect to the trap center the Coulomb potential is approximated by \cite{rudolph2022force},
\begin{eqnarray}
    V_{e} = -\frac{Q_{a}Q_{b}}{8\pi \epsilon_{0} d^{3}}(\mathbf{q}_{a}-\mathbf{q}_{b})^{2}
\end{eqnarray}
where $Q_{a,b}$ denotes the charges of the particles and $ d $ the interparticle separation. This leads to a change in the bare mechanical frequencies $ \omega_{a,b} $ given by 
\begin{eqnarray}
    \Omega_{a,b} = \sqrt{\omega_{a,b}^{2} - \frac{Q_{a}Q_{b}}{4\pi \epsilon_{0} m d^{3}}}
\end{eqnarray}
and a linear coupling with rate 
\begin{eqnarray}
    g_{e} = -\frac{Q_{a}Q_{b}}{4\pi \epsilon_{0} \hbar}\left( \frac{q_{0,a}q_{0,b}}{d^{3}} \right)
\end{eqnarray}
where $ q_{0,a}, q_{0,b}$ are the zero point fluctuations of each oscillator now defined in terms of $ \Omega_{a} $ and $ \Omega_{b} $, respectively. The sign of the interaction depends on the charges of the particles. Oppositely charged particles have $ g_{e} > 0 $, while like charged particles have $ g_{e} < 0 $. 
Note that the interaction can also be switched on and off by controlling the charge of the nanoparticles \cite{frimmer2017controlling,moore2021searching,ricci2022chemical}. This can be exploited to prepare the particle in an initial separable state and subsequently turn the interaction on over a time scale much shorter than $ \omega_{a}^{-1}, \omega_{b}^{-1} $.
Throughout this section, we will consider the parameters in Table \ref{parameter_table_coulomb} with values adapted from \cite{rieser2022tunable, rudolph2022force}.

\begin{table}[t!]
\begin{tabular}{cccc}
\hline \hline
Parameter  &   Symbol    & Units & Value \\ \hline 
 Charge &  $ Q_{a,b} $ &   $e$    &  $ 250 $ \\
 Interparticle distance &  $ d $ &   \SI{}{\mu m}    &  $ 2.0 $ \\
 Bare frequency $ a $ &  $ \omega_{a} $ &   \SI{}{kHz}    &  $ 2\pi \times 190$ \\
 Bare frequency $ b $ &  $ \omega_{b} $ &   \SI{}{kHz}    &  $ 2\pi \times 180$ \\
 Modified frequency $ a $ &  $ \Omega_{a} $ &   \SI{}{kHz}    &  $ 2\pi \times 147$ \\
 Modified frequency $ b $ &  $ \Omega_{b} $ &   \SI{}{kHz}    &  $ 2\pi \times 134$ \\
 Zero point fluctuation $ a $ &  $ q_{0,a} $ &   \SI{}{m}    &  $ 4.1 \times 10^{-12} $ \\
 Zero point fluctuation $ b $ &  $ q_{0,b} $ &   \SI{}{m}    &  $ 4.3 \times 10^{-12}$ \\
 Coulomb coupling rate &  $ g_{e} $ &   \SI{}{kHz}    &  $ 2\pi \times 51 $ \\
 Coupling-to-freq. ratio &  $ g_{e}/\Omega_{a,b} $ &   -   &  $ \approx 0.34 $ \\

    \hline \hline 
\end{tabular}
\caption{\label{parameter_table_coulomb} Coulomb interaction parameters. Values adapted from \cite{rieser2022tunable, rudolph2022force}.}
\end{table}

\begin{figure}[t!]
    \centering
    \includegraphics[width=0.5\textwidth]{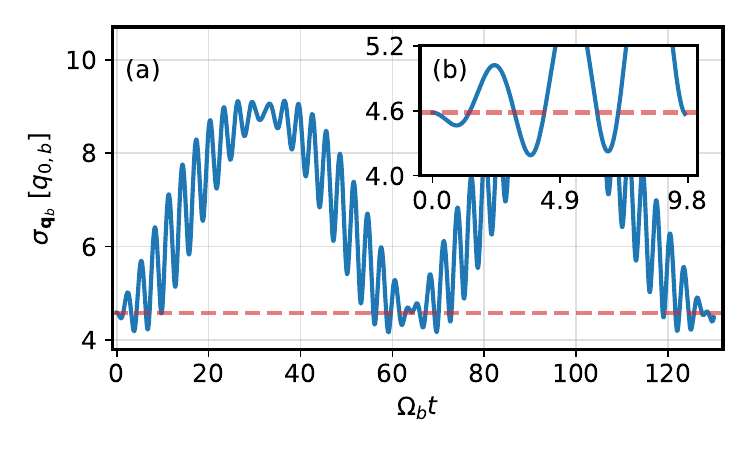}
    \caption{(a) Position uncertainty of a thermal semiclassical particle with $ \Bar{n}_{b} = 10 $ phonons influenced by a quantum particle in the ground state via Coulomb interaction (blue curve), in comparison to the thermal state position standard deviation $ \sigma_{0} \approx 4.6q_{0,b} $ (red dashed line). (b) Inset: repeated cooling and heating of the semiclassical particle.}
    \label{fig:pp-vacuum}
\end{figure}

\subsection{Quantitative estimations}

\begin{figure*}[ht!]
    \centering
    \includegraphics[width=\textwidth]{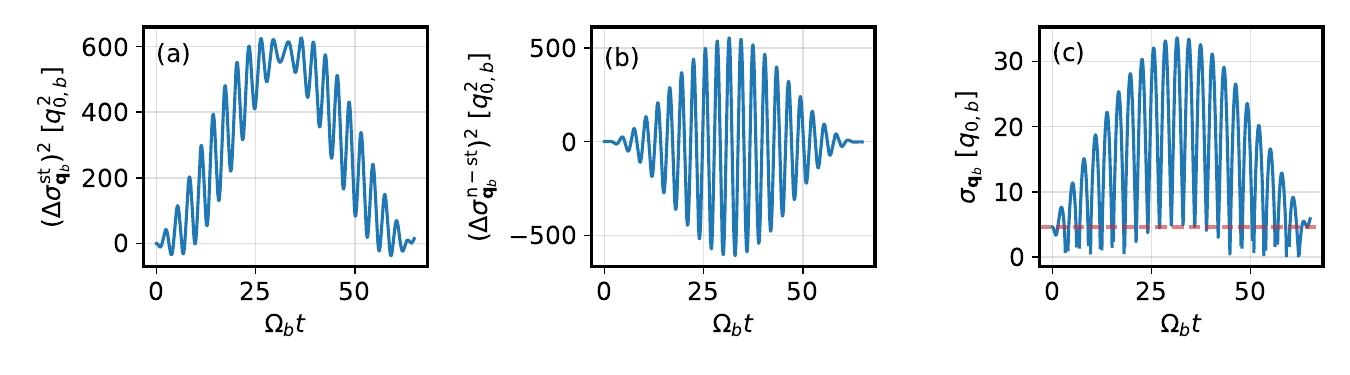}
    \caption{Effect of a squeezed quantum particle with $ r = 3 $ ($\approx 30$ dB) in contact with a semiclassical particle in an initially thermal state with occupation number $ \bar{n}_{b} = 10 $: (a) stationary contribution, (b) non-stationary contribution for a squeezing phase $\phi = 0 $ (solid blue curve), (c) total position rms for $\phi = 0 $ (solid blue curve) compared to the initial uncertainty $ \sigma_{0} = \sqrt{2\bar{n}_{b} + 1} \times q_{0,b} \approx 4.6 \times q_{0,b}  $ (red dashed line).}
    \label{fig:pp-squeezing}
\end{figure*}

Following Secs. \ref{trapedparticle} and \ref{sec:light-particle} we can write the excess rms of the semiclassical particle linearly coupled to the quantum particle in the ground state,
\begin{eqnarray}
    \Delta \sigma_{\mathbf{q}_{b}}^{2} = \frac{1}{(\kappa^2 - 1)^{2}}\left( \frac{\mathbf{f}_{0}}{m\Omega_{b}^2}   \right)^{2} h(t)
\end{eqnarray}
where $ h(t) $ is given in Eq. \eqref{oscillations1} and we have redefined $ \kappa = \Omega_{a} / \Omega_{b}$. Here the characteristic force $ \mathbf{f}_{0} $ defined in Eq. \eqref{characteristic_force} is evaluated at the Coulomb rate $ g_{e}$. For the parameters in Table \ref{parameter_table_coulomb}, $ \mathbf{f}_{0} \approx 7\times 10^{-18} $ N. 
Figure \ref{fig:pp-vacuum}(a) shows the position rms, given by Eq.~\eqref{excessrms}, of the semiclassical particle as a function of time, initially in a thermal state with mean occupation number $ \Bar{n}_{b} = 10 $ and standard deviation $ \sigma_{0} = \sqrt{2\Bar{n}_{b} + 1} \times q_{0,b} \approx 4.6 \times q_{0,b} $. As a consequence of the interaction, the standard deviation cyclically oscillates between $ \sigma_{0} $ and $ \approx 9\times  q_{0,b} $, representing heating and recooling of the particle motion. Note the position rms can become smaller than $ \sigma_{0} $, as shown in the inset in Figure \ref{fig:pp-vacuum}(b). 

A squeezed mechanical state will yield an exponentially enhanced effect, with the excess stationary quantum noise given by,
\begin{equation}  (\Delta\sigma_{\mathbf{q}_{b}}^{\rm st})^{2}= \cosh{(2r)}\Delta \sigma_{\mathbf{q}_{b}}^{2}
\end{equation}
and a non-stationary contribution
\begin{equation}
(\Delta\sigma_{\mathbf{q}_{b}}^{\rm n-st})^{2} =  \frac{\sinh(2r)}{( \kappa^{2}-1)^{2}}\left( \frac{\mathbf{f}_{0}}{m\Omega_{b}^2}   \right)^{2} h_{\phi}(t)\label{nstatvar}
\end{equation}
where $ h_{\phi}(t) $ carries the information on the squeezing phase and is defined as
\begin{align}
    h_{\phi}(t) =\cos (2 (\phi - \Omega_a t)) +\cos ( \Omega_b t) (\cos (2 \phi ) \cos ( \Omega_b t) \nonumber\\ 
    -2 \cos (2 \phi - \Omega_a t)) + 2\kappa \sin ( \Omega_bt) (\sin (2 \phi ) \cos ( \Omega_bt) \nonumber\\
    -\sin (2 \phi -\Omega_a t)) -\kappa^2 \cos (2 \phi ) \sin ^2( \Omega_b t)\label{difphases}
\end{align}
Figure \ref{fig:pp-squeezing} shows the individual stationary (a) and non-stationary (b) contributions and the total (c) position rms for a semiclassical particle in a thermal state with occupation number $ \bar{n}_{b} = 10 $ in contact with a squeezed quantum particle with squeezing parameter $ r = 3 $ ($\approx 30$ dB) \cite{kustura2022mechanical}, for  $\phi = 0 $, and for different angles in the supplemental material. The effect of different angles can be calculated from Eq.~\eqref{difphases} and Eq.~\eqref{nstatvar}. The individual stationary and non-stationary contributions can become negative, but observe the total position rms remains positive due to the thermal bath contribution. Moreover, the total position rms $\sigma_{\mathbf{q}_{b}}$ depends strongly on the squeezing phase, as can be seen from the traces in Figure \ref{fig:pp-squeezing}(c); see Supplementary video for a complete sweep of the squeezing phase from $ \phi = 0 $ to $ \phi = 2\pi $. 

\begin{figure}[t!]
    \centering
    \includegraphics[width=0.5\textwidth]{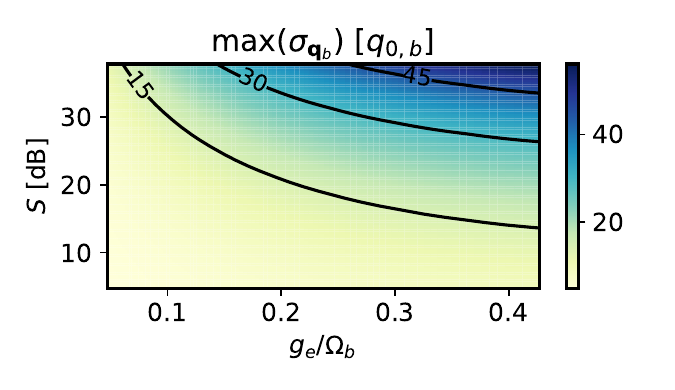}
    \caption{Dependence of the maximum value of the position rms $ \sigma_{\mathbf{q}_{b}} $ with the interaction coupling strength $ g_{e} / \Omega_{b}$ and quantum particle squeezing $ S $ (dB). The position rms is given in units of zero point fluctuations $ q_{0,b} $, and should be compared to the initial semiclassical uncertainty of $ \sigma_{0} \approx 4.6 \times q_{0,b}  $. The coupling strength is changed by varying the electric charge of the particles from $100$ to $ 260$ elementary charges.}
    \label{fig:pp-coupling}
\end{figure}

\begin{figure}[t!]
    \centering
    \includegraphics[width=0.48\textwidth]{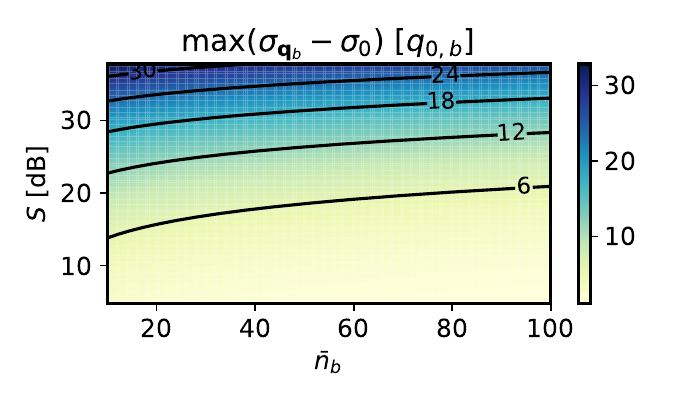}
    \caption{Dependence of the maximum value of the position rms $ \sigma_{\mathbf{q}_{b}} $ (in units of zero point fluctuations $ q_{0,b} $) with the initial number of phonons $ n_{b} $ in the semiclassical oscillator versus quantum particle squeezing $ S $ (dB), compared to the initial semiclassical uncertainty of $ \sigma_{0} \approx 4.6 \times q_{0,b}  $. The coupling strength is assumed to be $ g_{e}/\Omega_{b} = 0.2 $.}
    \label{fig:pp-temp}
\end{figure}

Figure \ref{fig:pp-coupling} shows the maximum value of $ \sigma_{\mathbf{q}_{b}}$ as a function of squeezing and the Coulomb coupling strength, changed by varying the particles' charge. Squeezing is measured in decibels (dB) by $ S = 10\log(2 \mathrm{Var}(q_{b})) $, with $ \mathrm{Var}(q_{b}) = e^{2r}/2 $. We see that for a quantum particle with 30 dB squeezing \cite{kustura2022mechanical} and moderate values of the coupling strength, $ g_{e} / \Omega_{b} \approx 0.2  $, significant enhancement of the position rms can be achieved when compared to the initial semiclassical rms value $ \sigma_{0} \approx 4.6 \times q_{0,b}  $. In terms of thermal occupation number, this represents an increase from an initial value of $ \bar{n}_{b} = 10 $ to $ \bar{n}_{b}' = 220 $, while for more modest 10 dB of squeezing, the increase of the occupation number will be from $ \bar{n}_{b} = 10 $ to $ \bar{n}_{b}' = 71 $. A possible way of measuring this effect of quantum-induced noise would be to employ a generalization of the Kalman filter as used in \cite{magrini2021real} to achieve zero point fluctuation-level position uncertainty in the presence of colored noise \cite{simon2006optimal}. For comparison, Figure \ref{fig:pp-temp} displays the maximum position rms $ \sigma_{\mathbf{q}_{b}} $ compared to $ \sigma_{0} $ as a function of the number of phonons of the semiclassical oscillator and squeezing of the quantum particle, at a fixed optomechanical coupling of $g_{e}/\Omega_{b} = 0.2$; as expected, the higher the number of phonons (the initial temperature of the semiclassical particle), the harder it becomes to observe the rms oscillations, unless more squeezing is added to the quantum oscillator.
For a single mode squeezed thermal state the excess position rms acquires an additional enhancement factor,
\begin{equation}
    (\Delta\sigma_{\mathbf{q}_{b}}^{\rm st})_{\mathrm{th}}^{2}= \coth\left(\frac{\beta_a \Omega_a \hbar}{2}\right)(\Delta\sigma_{\mathbf{q}_{b}}^{\rm st})^2,
\end{equation}
which becomes significant when $\beta_a^{-1}\gg \hbar \Omega_a/2$. Considering a levitated nanoparticle cooled to a number of phonons of $n_b=0.5$ ($T_a= 45\,$mK, $\beta_a =1.6 \times 10^{27}\,$ J/K) \cite{magrini2021real} we have $\coth\left(\beta_a  \hbar \Omega_a/2\right)\sim 2.12$, further enhancing the fluctuations in the semiclassical particle.


\subsection{Gravitational field of a delocalized particle}

A this point, we cannot resist exploiting the analogy between the Coulomb and Newtonian potentials to draw some comments on the stochastic gravitational field of a delocalized quantum particle.

We consider two identical particles of mass $ m $ at a center-of-mass separation $ d $. To leading order in the particles' displacements, the Newtonian gravitational potential yields an interaction Hamiltonian of the form \cite{krisnanda2020observable} $ H_N \approx (Gm^{2}/d^{3}) (\mathbf{q}_{a} - \mathbf{q}_{b})^{2}$, which translates into an effective frequency shift of
\begin{eqnarray}
    \Omega_{a,b} \approx \sqrt{\omega_{a,b}^{2} + 2Gm/d^{3} }
\end{eqnarray}
and a Newtonian coupling rate 
\begin{eqnarray}
    g_{N} \approx - \frac{2G}{\hbar} \frac{m^{2}q_{0,a}q_{0,b}}{d^{3}}
\end{eqnarray}
According to Eqs. \eqref{stationary_squeezed} and \eqref{nonstationary_squeezed}, for large squeezing parameters, we can recast both the stationary and non-stationary stochastic forces according to
\begin{eqnarray}
    \mathbf{f}_{\mathcal{Q}}(t) \approx e^{r} \frac{\hbar g_{N}}{q_{0,b}} \xi(t) = \frac{\hbar \Gamma_{\mathrm{ent}}}{q_{0,b}} \xi(t)
    \label{quantum_force_entanglement}
\end{eqnarray}
where $ \xi(t) $ satisfies either the stationary or non-stationary correlators in Eqs. \eqref{dimensionless_correlators1} and \eqref{dimensionless_correlators2} and we define the \textit{entanglement rate}
\begin{eqnarray}
    \Gamma_{\mathrm{ent}} \equiv 2\left(\frac{G}{\hbar}\right)\left(  \frac{m^{2} \Delta \mathbf{q}_{a} \Delta \mathbf{q}_{b}}{d^{3}} \right)
    \label{entanglement_rate}
\end{eqnarray}
where $ \Delta \mathbf{q}_{a} = e^{r} q_{0,a} $ and $ \Delta \mathbf{q}_{b} = q_{0,b} $ are the wavefunction uncertainties in the position basis for a squeezed particle and a particle in the ground state. 

To arrive at the stochastic force, we have considered a stationary phase approximation in the stochastic propagator in Eq. \eqref{final__mechanical_propagator}. If we do not make the semiclassical approximation, we can see that the gravitational interaction generates entanglement between the oscillators \cite{weiss2021large, krisnanda2020observable,al2018optomechanical, bengyat2023gravity} at a rate given by $ \Gamma_{\mathrm{ent}}  $ -- indeed Eq.  \eqref{entanglement_rate} is the short-time approximation of the entanglement rate between two continuously delocalized oscillators interacting via gravity as derived in \cite{bengyat2023gravity}. 
Arguably, the rate $ \Gamma_{\mathrm{ent}}  $ is extremely small given current experiments, but the result  \eqref{quantum_force_entanglement} is conceptually interesting as it makes the connection between entanglement and the quantum-induced stochastic force of a subsystem manifest. 
Naturally, the same conclusions apply to the much stronger case of Coulomb interactions. 


\section{Discussion \label{sec:dissucssion}}

In this work, we have applied the formalism of double path integrals and Feynman-Vernon influence functionals to linear optomechanical systems. We have analysed the effective stochastic dynamics induced by the interactions between a semiclassical and a quantum system in different states, notably the ground state, squeezed-coherent and squeezed-thermal states. Colored noise and dissipation with memory arising from quantum fluctuations are ubiquitous in linear optomechanical quantum-classical interactions. Microscopically, these fluctuations can be understood as a `semiclassical' manifestation of the entanglement generated by the interaction between the two subsystems.
We have studied these effects in the context of levitated nanoparticles, both in cavity and free space multi-particle scenarios. Notably, the quantum-classical stochastic dynamics induced by the Coulomb interaction between two levitated particles is potentially measurable in near future experiments with delocalized quantum states, where squeezing of the mechanical wavefunction yields an exponential enhancement of the quantum-induced stochastic forces. The analogy between the Coulomb and Newtonian potentials has been used to comment on the connection between the effective stochastic dynamics and gravitational-induced entanglement. 

Throughout our study, we find strong similarities between a linear optomechanical system and the effective field theory of a quantized gravitational wave mode interacting with a GW detector, despite both systems being governed by seemingly different Hamiltonians, one originating from the Einstein-Hilbert action \cite{parikh2021signatures} while the other from the interaction of dielectrics in electromagnetic fields \cite{gonzalez2019theory, cheung2012optomechanical}. This reinforces the analogies between optomechanics and the quantum theory of GWs \cite{guerreiro2020quantum} and opens the possibility of investigating novel ways of probing the quantum nature of GWs through proof-of-principle laboratory experiments. 

Our findings also open the way to novel investigations at the interface between stochastic thermodynamics and fundamental tests of quantum theory. For instance, the formalism of path integrals allows for the calculation of the work distribution \cite{paraguassu2022effects} and probability of rare violations of the second law of thermodynamics \cite{paraguassu2022probabilities} springing from state dependent quantum-induced fluctuations in levitated optomechanical systems. Moreover, the formalism can be extended to investigate the effective stochastic dynamics induced by multipartite entangled quantum states in contact with a classical probe. In principle, such investigation could lead to new forms of witnessing entanglement and non-classicality in the mesoscopic scale.

\section*{Acknowledgements}
We acknowledge Carlos Tomei, Nicolau Saldanha, Welles Morgado, Bruno Suassuna, Oscar Kremer, Ariel Hertz for useful discussions. 
P.V.P acknowledges the Funda\c{c}\~ao de Amparo \`a Pesquisa do Estado do Rio de Janeiro (FAPERJ Process SEI-260003/000174/2024). T.G. acknowledges the Coordena\c{c}\~ao de Aperfei\c{c}oamento de Pessoal de N\'ivel Superior - Brasil (CAPES) - Finance Code 001, Conselho Nacional de Desenvolvimento Cient\'ifico e Tecnol\'ogico (CNPq), Funda\c{c}\~ao de Amparo \`a Pesquisa do Estado do Rio de Janeiro (FAPERJ Scholarship No. E-26/200.252/2023 and E-26/202.762/2024)  and Funda\c{c}\~ao de Amparo \`a Pesquisa do Estado de São Paulo (FAPESP process No. 2021/06736-5). This work was supported by the Serrapilheira Institute
(grant No. Serra – 2211-42299 ) and StoneLab. 

\section*{Disclosures}The authors declare that there are no conflicts of interest
related to this paper.

\bibliography{main}

\newpage


\appendix

\onecolumngrid

\section{Fluctuation and Dissipation in a cavity}\label{integralsapp}

Summation over modes in a cavity leads us to the fluctuation and dissipation integrals in Eqs.  \eqref{stat_noise} and \eqref{non_stat_noise}. Taking into account the density of modes in Eq. \eqref{density} results in the two main integrals,
\begin{eqnarray}
    I_{1} = \int_{0}^{\infty} d\omega \frac{\omega^4\cos(\omega(t-t'))}{(\omega-\omega_c)^2+\gamma^2} \\
    I_{2} = \int_{0}^{\infty} d\omega \frac{\omega^4\sin(\omega(t-t'))}{(\omega-\omega_c)^2+\gamma^2}
\end{eqnarray}
for the fluctuation and dissipation terms, respectively.  
These can be rewritten as
\begin{eqnarray}
    I_{1} = \frac{d^4}{d\tau'^4}J_{1} ,\label{integraltrick1} \\
    I_{2} = \frac{d^4}{d\tau'^4}J_{2}.\label{integraltrick2}
\end{eqnarray}
where, 
\begin{eqnarray}
    J_{1} = \int_0^\infty d\omega \frac{\cos(\omega\tau) }{(\omega-\omega_c)^2+\gamma^2} \\
    J_{2} = \int_0^\infty d\omega \frac{\sin(\omega\tau)}{(\omega-\omega_c)^2+\gamma^2}\label{J2}
\end{eqnarray}
and $\tau=t-t'$. Both $ I_{1} $ and $ I_{2} $ can be evaluated in terms of distributions. We now proceed to calculate each integral, starting with $ J_{1,2}$ and then $ I_{1,2} $.

\subsection{Evaluation of $J_{1,2}$}

 {Eq. \eqref{J2}} can be written in terms of a Fourier transform,
\begin{equation}
  \int_0^\infty d\omega \frac{ \sin{\omega \tau}}{\left(\omega - \omega_c\right)^2+\gamma^2}  = \mathrm{Im} \int_{-\infty}^\infty d\omega \frac{ H(\omega) e^{i\omega \tau}}{\left(\omega - \omega_c\right)^2+\gamma^2},\label{fourier_transform_convolution1}
\end{equation}
We can use the convolution theorem to solve \eqref{fourier_transform_convolution1}. We use the following convention for the convolution of two functions,
\begin{equation}
    \frac{1}{2\pi}(f*g)(\tau) =\frac{1}{2\pi} \int_{-\infty}^\infty d\alpha f(\alpha)g(\tau-\alpha) .\label{conv}
\end{equation}
For the Heaviside, we have the Fourier transform,
\begin{equation}
    f(\tau) =  \int_{-\infty}^\infty d\omega H(\omega) e^{i\omega \tau}  = \pi \delta(\tau) + \mathcal{P}\left(\frac{i}{\tau}\right),
\end{equation}
where $\mathcal{P}$ denotes the Cauchy principal value. Using \cite{zee2010quantum},
\begin{equation}
    \mathcal{P}\left(\frac{1}{\tau}\right) = \frac{1}{\tau + i\epsilon} + \pi i \delta(\tau),
\end{equation}
we have
\begin{equation}
    f(\tau) =  \int_{-\infty}^\infty d\omega H(\omega) e^{i\omega \tau}  = \frac{i}{\tau + i\epsilon}.
\end{equation}
For the Lorentzian the Fourier transform reads,
\begin{equation}
    g(\tau) = \int_{-\infty}^\infty d\omega \frac{  e^{i\omega \tau}}{\left(\omega - \omega_c\right)^2+\gamma^2}  = \frac{\pi}{\gamma} e^{i\omega_c\tau}e^{-\gamma |\tau|} \label{intdiss}.
\end{equation}

Using the convolution theorem,
\begin{equation}
   \frac{1}{2\pi}(f*g)(\tau)= \frac{i}{2\gamma}\int_{-\infty}^\infty d\alpha \frac{1}{\tau-\alpha+ i \epsilon}e^{i\omega_c\alpha}e^{-\gamma |\alpha|}. \label{int2}
\end{equation}
We can again use the convolution theorem to solve \eqref{int2}. We have,
\begin{equation}
    \int_{-\infty}^\infty d\beta e^{i\omega_c\beta}e^{-\gamma |\beta|} = \frac{2\gamma}{\gamma^2+\omega_c^2},\label{f1}
\end{equation}
\begin{equation}
    \int_{-\infty}^\infty d\beta \frac{1}{\tau-\beta\pm i \epsilon}e^{i\omega_c \beta}  = -2\pi i e^{i\omega_c\tau},\label{f2}
\end{equation}
where $ \omega_{c} > 0 $. 
The convolution of Eq.~\eqref{f1} and (\ref{f2}) reads,
\begin{equation}
  \int_{-\infty}^\infty d\beta \frac{e^{i\omega_c\beta}e^{-\gamma |\beta|}}{\tau-\beta\pm i \epsilon} = -2 i\gamma e^{i\omega_c\tau}\int_{-\infty}^{\infty} d\alpha \frac{ e^{-i\tau\alpha}}{\gamma^2+\alpha^2}. 
\end{equation}
Finally, using
\begin{equation}
    \int_{-\infty}^{\infty} d\alpha \frac{ e^{-i\tau\alpha}}{\gamma^2+\alpha^2}  = \frac{\pi e^{-\gamma |\tau|}}{\gamma},
\end{equation}
we find,
\begin{equation}
   \frac{i}{2\gamma} \int_{-\infty}^\infty d\alpha \frac{e^{i\omega_c\alpha}e^{-\gamma |\alpha|}}{\tau-\alpha\pm i \epsilon}  = \frac{\pi}{\gamma} e^{i\omega_c\tau}e^{-\gamma |\tau|}.
\end{equation}
Putting it all together and taking the imaginary part of the result we arrive at 
\begin{equation}
    \int_0^\infty d\omega \frac{ \sin{\omega \tau}}{\left(\omega - \omega_c\right)^2+\gamma^2}  = \frac{\pi}{\gamma}e^{-\gamma |\tau|}\sin\left(\omega_c\tau\right).
\end{equation}

The integral $ J_{1} $ can be written as a Fourier transform,
\begin{equation}
    J_{1} = \frac{1}{2} \mathrm{Re} \int_{-\infty}^\infty d\omega  \frac{  e^{i\omega \tau}}{\left(\omega - \omega_c\right)^2+\gamma^2} 
\end{equation}
Again using \eqref{intdiss} we have,
\begin{eqnarray}
    J_{1} = \frac{\pi}{2\gamma}e^{-\gamma|\tau|}\cos\omega_c\tau.
\end{eqnarray}

\subsection{Distributional derivatives}

We now need to take derivatives of $J_{1,2}$. Note these are discontinuous and exist only in the sense of distributions \cite{appel2007mathematics}. 
Define the action of a distribution $ T $ on a test function $ \phi $ as the inner product
\begin{eqnarray}
    \langle T , \phi \rangle = \int_{-\infty}^{\infty} dx T(x)\phi(x)
\end{eqnarray}
and the derivative of a distribution as
\begin{eqnarray}
     \langle T' , \phi \rangle = -  \langle T , \phi' \rangle 
\end{eqnarray}

\noindent To evaluate the derivatives of $J_{1,2}$ we note that discontinuities spring from the the $ e^{-\gamma \vert \tau \vert} $ terms, for which the first derivative is
\begin{equation}
    \frac{d}{d\tau}e^{-\gamma |\tau|} = -\gamma\, \rm{sign}({\tau})e^{-\gamma |\tau|}.
\end{equation}
We can evaluate higher derivatives by considering their effects on a test function $ \phi(\tau) $. We have,
\begin{eqnarray}
    \left\langle \frac{d}{d\tau}\left( -\gamma\, \rm{sign}({\tau})e^{-\gamma |\tau|}\right),\phi(\tau)\right\rangle \nonumber \\ = -\left\langle \left(-\gamma\, \rm{sign}({\tau}) e^{-\gamma |\tau|}\right), \phi'(\tau) \right\rangle \nonumber\\= \gamma \int_0^\infty e^{-\gamma \tau}\phi'(\tau)d\tau - \gamma\int_{-\infty}^0 e^{\gamma \tau}\phi'(\tau)d\tau \nonumber \\
    = -2\gamma \phi(0) - \gamma\int_{-\infty}^{\infty} \frac{d}{d\tau}\left(e^{-\gamma |\tau|}\right)\phi(\tau) d\tau \nonumber \\
    = -2\gamma \langle \delta(\tau),\phi(\tau)\rangle +\gamma^2 \langle e^{-\gamma |\tau|},\phi(\tau)\rangle.
\end{eqnarray}
 We conclude, 
\begin{equation}
     \frac{d^2}{d\tau^2}\left( e^{-\gamma |\tau|}\right) = -2\gamma\delta(\tau) +\gamma^2 e^{-\gamma |\tau|}.
\end{equation}
 {Note that in order to derive this equation, we considered the $\rm{sign}(\tau)$ function. Since $\tau$ represents a time interval, it can be either negative or positive. This fact is utilized when we split the integral into different intervals, using the property of the $\rm{sign}(\tau)$ function to introduce a negative sign in the third line of the equation above.}

Generalizing to higher orders we find,
\begin{eqnarray}
    \frac{d^3}{d\tau^3}\left( e^{-\gamma |\tau|}\right) = -2\gamma\delta'(\tau)-\gamma^3\rm{sign}({\tau})e^{-\gamma |\tau|},\\
    \frac{d^4}{d\tau^4}\left( e^{-\gamma |\tau|}\right) =-2\gamma\delta''(\tau)-2\gamma^3\delta(\tau) +\gamma^4 e^{-\gamma |\tau|}.
\end{eqnarray}

Through repeated applications of the above derivative rules we can compute Eqs.~\eqref{integraltrick1} and (\ref{integraltrick2}). We have,
\begin{eqnarray}
     \frac{d^4}{d\tau^4}\left(e^{-\gamma |\tau|}\sin(\omega_c\tau)\right) = \omega_{c}^{4} \left[  -2(\nu^{3} - 6\nu) \dfrac{\delta(\tau)}{\omega_{c}} \sin(\omega_{c}\tau) - 8 \nu \frac{\delta'(\tau)}{\omega_{c}^{2}} \cos(\omega_{c}\tau) - 2\nu \frac{\delta''(\tau)}{\omega_{c}^{3}} \sin(\omega_{c}\tau) \right. \nonumber \\
     \left. + e^{-\gamma\vert \tau'\vert} \left(   (1 - 6\nu^{2} + \nu^{4}) \sin(\omega_{c}\tau) + 4 (\nu - \nu^{3})\mathrm{sgn}(\tau) \cos(\omega_{c}\tau) \right) \right]
\end{eqnarray} 

\begin{eqnarray}
   \frac{d^4}{d\tau^4}\left(e^{-\gamma |\tau|}\cos(\omega_c\tau)\right) = \omega_{c}^{4} \left[  -2(\nu^{3} - 6\nu) \dfrac{\delta(\tau)}{\omega_{c}} \cos(\omega_{c}\tau)  {+} 8 \nu \frac{\delta'(\tau)}{\omega_{c}^{2}} \sin(\omega_{c}\tau) - 2\nu \frac{\delta''(\tau)}{\omega_{c}^{3}} \cos(\omega_{c}\tau) \right. \nonumber \\
     \left. + e^{-\gamma\vert \tau \vert} \left(   (1 - 6\nu^{2} + \nu^{4}) \cos(\omega_{c}\tau) - 4 (\nu - \nu^{3})\mathrm{sgn}(\tau) \sin(\omega_{c}\vert \tau \vert) \right) \right]
    \label{cfluc}
    \end{eqnarray}
where as in the main text we define $ \nu = \gamma/\omega_{c} $. We can further simplify this result by introducing certain distribution identities.

\subsection{Distribution identities}

Consider the action of the $ \delta''(\tau)\cos(\omega_c\tau) $ distribution on a test function,
 \begin{eqnarray}
     \langle \delta''(\tau)\cos(\omega_c\tau) ,\phi(\tau) \rangle &=& \langle \delta''(\tau) ,\cos(\omega_c\tau)\phi(\tau) \rangle  = \langle \delta(\tau) ,(\cos(\omega_c\tau)\phi''(\tau)) \rangle \nonumber \\
     &=& \langle \delta(\tau) ,\left(-2 \omega_c \sin (\tau  \omega_c) \phi '(\tau )+\cos (\tau  \omega_c) \phi ''(\tau )+\right.  \left.-\omega_c^2 \phi (\tau ) \cos (\tau  \omega_c)\right) \rangle \nonumber \\ &=& \phi''(0)-\omega_c^2\phi(0) = \langle \delta(\tau'')-\omega_c^2\delta(\tau) , \phi(\tau) \rangle
 \end{eqnarray}
Therefore, 
\begin{equation}
    \delta''(\tau)\cos(\omega_c\tau) = \delta''(\tau)-\omega_c^2\delta(\tau).
\end{equation}
Following similar steps, we also have the following identities, used throughout the main text,
\begin{eqnarray}
    \delta(\tau)\cos(\omega_c\tau) &=& \delta(\tau),\\
    \delta'(\tau)\sin(\omega_c\tau) &=& -\omega_c \delta(\tau), \\
    \delta'(\tau)\cos(\omega_c\tau) &=& -\delta'(\tau), \\
    \delta''(\tau)\sin(\omega_c\tau) &=& 2\omega_c\delta'(\tau). 
\end{eqnarray}

\subsection{Supplementary video}

The supplementary video shows the typical semiclassical particle's position rms $ \sigma_{\mathbf{q}_{b}}$ as a function of time, in units of zero-point fluctuations $q_{0,b}$ for different values of the squeezing phase $ \phi $ of the quantum oscillator (see Eq. (155) in the main text). For this plot, we assumed the semiclassical oscillator with an initial number of phonons $ n_{b} = 100 $ in contact with a quantum oscillator with squeezing parameter $r = 3$ ($\approx 30$ dB) and a coupling rate of $ g_{e}/\Omega_{b} = 0.2$.

\end{document}